\documentclass[manuscript,screen]{acmart}
\usepackage{multirow}
\usepackage[table]{xcolor}
\usepackage{pifont}
\usepackage{rotating, adjustbox}
\usepackage{comment}
\usepackage{paralist}

\usepackage{enumitem}
\usepackage{lmodern}
\usepackage{booktabs}
\usepackage{tabularx}
\newcolumntype{L}[1]{>{\raggedright\arraybackslash}p{#1}}
\usepackage{array}
\usepackage{booktabs}
\usepackage[table]{xcolor}
\usepackage{pifont} % For \ding
\usepackage{colortbl} % For \cellcolor
\usepackage{geometry}
\usepackage{makecell}
\usepackage{pgf-pie} % For pie charts
\usepackage{xcolor}  % For color customization
\usepackage[most]{tcolorbox} % for colored/shaded boxes
\AtBeginDocument{%
  }

\begin{document}

\title{Structuring Security: A Survey of Cybersecurity Ontologies, Semantic Log Processing, and LLMs Application}

\author{Bruno Lourenço}
\authornote{Both authors contributed equally to this research.}
\email{bruno.horta.lourenco@tecnico.ulisboa.pt}
\orcid{1234-5678-9000}
%\author{Pedro Adão}
%%\authornotemark[1]
%\email{pedro.adao@tecnico.ulisboa.pt}
\affiliation{%
  \institution{INESC-ID and CINAV, ULisboa and Portuguese Naval Academy}
  \city{Lisbon}
  %%\state{Ohio}
  \country{Portugal}
}

\author{Pedro Adão}
\affiliation{%
  \institution{Instituto Superior Técnico, ULisboa}
  \city{Lisbon}
  \country{Portugal}}
\email{pedro.adao@tecnico.ulisboa.pt}

\author{João F. Ferreira}
\affiliation{%
  \institution{INESC-ID and Faculty of Engineering, University of Porto}
  \city{Lisbon}
  \country{Portugal}}
\email{joao@joaoff.com}

\author{Mario Monteiro Marques}
\affiliation{%
  \institution{CINAV, Portuguese Naval Academy}
  \city{Lisbon}
  \country{Portugal}}
\email{mario.monteiro.marques@marinha.pt}

\author{Cátia Vaz}
\affiliation{%
  \institution{INESC-ID and Instituto Superior de Engenharia de Lisboa, Instituto Politécnico de Lisboa}
  \city{Lisbon}
  \country{Portugal}}
\email{cvaz@cc.isel.ipl.pt}

%\renewcommand{\shortauthors}{Lourenço et al.}

\begin{abstract}
This survey investigates how ontologies, semantic log processing, and Large Language Models (LLMs) enhance cybersecurity. Ontologies structure domain knowledge, enabling interoperability, data integration, and advanced threat analysis. Security logs, though critical, are often unstructured and complex. To address this, automated construction of Knowledge Graphs (KGs) from raw logs is emerging as a key strategy for organizing and reasoning over security data. LLMs enrich this process by providing contextual understanding and extracting insights from unstructured content. This work aligns with European Union (EU) efforts such as NIS 2 and the Cybersecurity Taxonomy, highlighting challenges and opportunities in intelligent ontology-driven cyber defense.
\end{abstract}



%\received{XX February XXXX}
%\received[revised]{XX March XXXX}
%\received[accepted]{XX June XXXX}

\maketitle

\section{INTRODUCTION}
Ontology-based approaches to information security have emerged as a pivotal research area in response to the growing complexity and sophistication of cyber threats~\cite{Oliveira2021}. As organizations increasingly rely on interconnected digital infrastructures, the need for structured, semantically rich security models becomes critical to enhance threat detection, reasoning, and informed decision-making.

This urgency is reinforced by recent regulatory developments, most notably the implementation of the NIS 2 Directive from the European Union (EU) 2022/2555~\cite{EU2022NIS2}, which came into effect on 16 January 2023. Replacing Directive EU 2016/1148, NIS 2 establishes stricter cybersecurity obligations for entities delivering essential or important services within the EU.

Ontologies, defined as formal, explicit specifications of shared conceptualization~\cite{Gruber1993}, offer a promising foundation to address these requirements. They model domain knowledge using fundamental constructs such as entities, relations, roles, and resources, enabling comprehensive and machine-interpretable representations of security domains~\cite{Moller2020}.

Semantic Web technologies are instrumental in the operationalization of these ontology-based frameworks. Standards such as the Resource Description Framework (RDF) and the Web Ontology Language (OWL) provide syntactic and semantic foundations for knowledge sharing, integration, and reuse across organizational and application boundaries.

Unlike traditional syntactic models, RDF and OWL facilitate the conceptual encoding of information through ontologies, which serve as the backbone of semantic Knowledge Graphs (KGs). This semantic foundation supports automated inference and reasoning, enhancing interoperability between heterogeneous systems and enabling context-aware analysis of cybersecurity events~\cite{Sikos2023}.

Representing knowledge in the form of KGs further expands analytical capabilities by integrating ontological structure with graph-based data models. Deductive reasoning within KGs is powered by rule-based semantics provided by RDF Schema (RDFS) and OWL, allowing for the logical entailment of new facts from existing knowledge bases~\cite{Polleres2013}.
This ability to infer implicit knowledge from explicitly defined relationships transforms static data into actionable intelligence, which is especially valuable in dynamic domains like cybersecurity.

Furthermore, graph query languages play a crucial role in enabling efficient retrieval and exploration of knowledge within KGs, thus supporting complex analytical workflows~\cite{Angles2017}. Recent advances highlight the growing synergy between Large Language Models (LLMs) and KGs. Although LLMs offer powerful generative and interpretive capabilities, KGs provide structured, context-rich representations, and ontologies serve to formalize domain knowledge, support reasoning, and enhance the explainability and reliability of LLM-driven systems. In the context of cybersecurity, the integration of LLMs holds promise for improving threat detection, automating vulnerability assessment, and enabling intelligent defense strategies. 

Another critical component of security monitoring is the system log files, which record events and activities across systems. However, their analysis is often challenged by high volume, heterogeneity, and structural complexity~\cite{Ekelhart2021}. Ontologies facilitate structured log analysis and enable automated reasoning, enhancing threat correlation and significantly reducing the need for manual inspection.

Cybersecurity KGs extend these benefits by providing graph-based representations of cyber-knowledge. They support visualization of attack paths, identification of data flows, and aggregation of intelligence from multiple sources~\cite{Sikos2023}. 

This paper builds on the foundations of ontologies, KGs, logs, and LLMs to explore how they can be combined to enhance and strengthen cybersecurity.
This approach aligns with the NIS 2 Directive, which underscores the need for automation, interoperability, and adaptability in cyber threat management. To support this alignment, the JRC Cybersecurity Taxonomy \cite{JRC2021CybersecurityTaxonomy} is employed to classify cybersecurity ontologies according to defined domains and subdomains.

This survey represents the first in-depth effort to collect the approaches that have being used in the domains and subdomains of the JRC Cybersecurity Taxonomy using ontologies, KGs and LLM, presenting the following key contributions: 
\begin{enumerate}[leftmargin=1.2cm, label=\textbf{\arabic*.}]
    \item \textbf{Novel Contribution to Cybersecurity Ontology Research}: This survey offers a focused investigation of the use of ontologies in the cybersecurity landscape, integrating principles from knowledge representation to Semantic Web technologies to address domain-specific challenges.

    \item \textbf{Comprehensive Review of Ontology-Based Cybersecurity Domains}: This survey presents the first systematic analysis of ontology applications across the full range of cybersecurity domains defined by the JRC Cybersecurity Taxonomy \cite{JRC2021CybersecurityTaxonomy}. For each subdomain, it compiles relevant cybersecurity ontologies that leverage semantic log extraction, often incorporating log ontologies and LLMs approaches, to enable intelligent inference from log data and enhance threat detection and overall cybersecurity effectiveness.    
\end{enumerate}

% 2 Review Scope and Methodology
\section{REVIEW SCOPE AND METHODOLOGY}
\label{sec:2}
Given the foundational role of ontologies in representing structured knowledge for cybersecurity and enabling semantic interoperability, this survey begins by examining the conceptual underpinnings and practical importance of ontologies within the Semantic Web and cybersecurity domains (Section \ref{sec:2.1}). Building on this foundation, we then define the scope of our analysis by selecting a representative subset of cybersecurity domains from the JRC Cybersecurity Taxonomy, focusing on those with strategic relevance, emerging research, and practical applicability (Section \ref{sec:2.2}). Finally, we describe the methodology adopted to identify, select, and analyze relevant literature, including our search strategy, inclusion criteria, and comparative framework for evaluating existing ontologies and their modeled content (Section \ref{sec:2.3}).

% 2.1 Ontologies
\subsection{Ontologies}
\label{sec:2.1}
Ontologies are fundamental Semantic Web technologies and are considered its backbone. They are used to represent knowledge within a specific domain by defining the concepts involved and the relationships between them. Similarly, they serve as a representational vocabulary, often specialized for a particular domain or subject matter~\cite{Chandrasekaran2002}.

In the field of cybersecurity, an ontology serves as a formal representation of cyberspace concepts and properties, structured through upper and domain ontologies. These ontologies capture the semantics of network topologies, devices, information flows, vulnerabilities, and cyberthreats, enabling application-specific, situation-aware querying and knowledge discovery through automated reasoning. 

OWL ontologies are divided into three sub-languages—OWL Lite, OWL DL, and OWL Full—ordered by increasing expressiveness, from the simpler OWL Lite, which supports basic hierarchies and constraints, to OWL DL, which balances expressiveness with computational completeness, and finally to the more comprehensive OWL Full, which provides maximum flexibility at the cost of decidable reasoning~\cite{McGuinness2003}. 

The effective use of ontologies requires not only a well-designed and well-defined ontology language, but also support from reasoning tools. Reasoning
is important both to ensure the quality of an ontology, and to exploit
the rich structure of ontologies and ontology based information~\cite{Staab2009}.

Ontologies play a central role in the realization of FAIR (findable, accessible, interoperable, and reusable) principles, particularly with respect to \emph{interoperability} and \emph{reusability}. FAIR guidelines emphasize the use of formal, accessible, shared, and broadly applicable vocabularies for knowledge representation. They also call for qualified references to other data or metadata and adherence to domain-specific community standards. In addition, ontologies contribute to \emph{findability} by enabling rich metadata descriptions and supporting \emph{accessibility} by ensuring that metadata can be retrieved via persistent and unique identifiers~\cite{Poveda2020}.

At the same time, there is growing consensus that the future of semantic technologies lies in their ability to support large-scale reasoning across complex, knowledge-intensive domains. These applications, often extending beyond the original scope of the Semantic Web, depend on expressive and high-quality ontologies. Enhancing ontology expressiveness, for example through the inclusion of disjointness axioms, can significantly improve reasoning capabilities and support automatic ontology evaluation. Such advances not only improve the reliability and coherence of ontologies but also further reinforce their critical role in achieving FAIR data compliance~\cite{Ciaramita2008}.

% 2.2 Cybersecurity Review Scope
\subsection{Cybersecurity Review Scope}
\label{sec:2.2}

This section presents the review scope and explains the process of searching, selecting, and analyzing the studies. We focus on identifying and analyzing existing ontologies within selected domains and subdomains from the EC JRC Cybersecurity Taxonomy \cite{JRC2021CybersecurityTaxonomy}.

Based on this taxonomy, 8 out of 15 domains were selected, each offering distinct contributions to the broader cybersecurity landscape, as detailed: \begin{inparaenum}[(\bgroup\bfseries 1\egroup)]
\item \textbf{Assurance, Audit, and Certification} ensures systems meet defined security objectives through formal evaluation methods; 
\item \textbf{Data Security and Privacy} protects data integrity, confidentiality, and privacy, ideally by design; 
\item \textbf{Education and Training} builds cybersecurity knowledge, skills, and workforce readiness; 
\item \textbf{Incident Handling and Digital Forensics} addresses digital evidence collection, analysis, and threat response; 
\item \textbf{Network and Distributed Systems} secures communication, infrastructure, and coordination in connected environments;

%###################

\item \textbf{Security Management and Governance} defines processes and policies ensuring core security properties and accountability; 

%###################

\item \textbf{Software and Hardware Security Engineering} integrates security across development life cycles; 
and \item \textbf{Theoretical Foundations} applies formal methods to rigorously prove security properties.
\end{inparaenum}

The selection of these 8 cybersecurity domains followed strict inclusion/exclusion criteria to ensure relevance, depth, and applicability. \emph{Assurance, Audit, and Certification} was included for its role in risk management and compliance. \emph{Security Management and Governance} shapes policies and risk frameworks, justifying its selection. \emph{Data Security and Privacy} has cross-sector importance, supporting privacy and regulatory needs.
\emph{Education and Training} supports workforce development and talent cultivation.
\emph{Incident Handling and Digital Forensics} addresses operational readiness and threat response. \emph{Software and Hardware Security Engineering} ensures secure-by-design system development. \emph{Network and Distributed Systems Security} protects critical infrastructure and services, and \emph{Theoretical Foundations} offers formal, provable security and methodological rigor.

Excluded domains lacked novelty, broad relevance, or strategic focus. \emph{Cryptology} and \emph{Identity and Access Management} were deemed mature and integration-focused.
\emph{Steganography} and related areas had niche or limited enterprise value. \emph{Human Aspects} and \emph{Legal Aspects} were seen as peripheral or externally governed and \emph{Trust Management and Accountability} and \emph{Security Measurements} were redundant within broader domains.

\subsection{Methodology}
\label{sec:2.3}
\begin{figure}[h]
    \centering
    \includegraphics[width=1.0\textwidth]{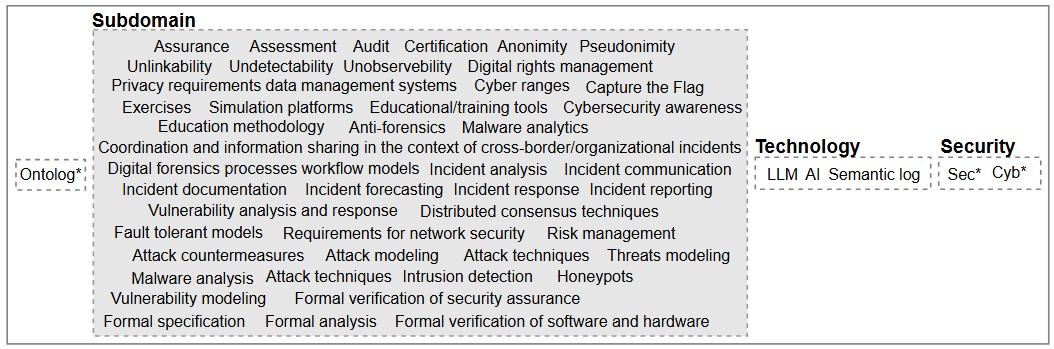}
    \caption{ Conceptual mapping of keywords used to construct the literature search strings, organized by subdomains, technology-related terms, and security-related terms.}
    \label{fig:ss1}
\end{figure}

To identify, select, and analyze the most recent and widely adopted ontologies in cybersecurity and its subdomains, and to assess whether they incorporate semantic log extraction or LLM-based technologies, we followed a three-step review process:
\begin{inparaenum}[(\bgroup\bfseries 1\egroup)]
\item identification of keywords covering cybersecurity ontologies, semantic log extraction, and LLMs; 
\item selection of relevant articles; and 
\item development of a comparative analysis based on a proposed classification of the information modeled in these ontologies.
\end{inparaenum}
As an initial step, we performed a Google Scholar search using keyword combinations that capture the core concepts of each cybersecurity subdomain. These search strings were executed either automatically, through a Python script connecting to the publisher’s API, or manually, following the query structure illustrated in Figure \ref{fig:ss1}.

For example, in the \emph{Assurance, Audit, and Certification} domain—which encompasses the subdomains \emph{Assurance, Assessment, Audit, and Certification}—we construct targeted queries to retrieve papers for the \emph{Assurance} subdomain. The combined search strings used for this purpose are shown in Figure \ref{fig:ss2}.

\begin{figure}[h]
\centering
\definecolor{lightgray}{RGB}{240,240,240}
%\definecolor{lightorange}{RGB}{255,200,150}
\begin{tcolorbox}[
  colback=lightgray,
  colframe=lightgray,
  boxrule=0pt,
  arc=0mm,
  left=2mm, right=2mm,
  top=1mm, bottom=1mm
]
\textbf{Search query 1:} Ontolog* AND <Subdomain>  

\textbf{Search query 2:} Ontolog* AND <Subdomain> AND (Secur* OR Cyb*)  

\textbf{Search query 3:} Ontolog* AND <Subdomain> AND (LLM OR AI OR Semantic Log) AND (Secur* OR Cyb*)  
\end{tcolorbox}
\caption{Structured search query formulations combining ontology-related terms with subdomain, technology, and security keywords for literature retrieval.}
\label{fig:ss2}
\end{figure}

\textbf{Note on Wildcard Usage}: To enhance replicability and transparency,
we utilized the wildcard character * in our search terms.
This allows the search engine to include all variations of a rootword,
thereby broadening the scope of our literature search. For instance,
“secur*” captures terms such as “security”, “secure”, “securing”, etc.,
ensuring that relevant studies using different terminologies are not
inadvertently excluded. In addition, we performed similar keyword searches on other academic databases, including Web of Science, the ACM Digital Library, and IEEE Xplore.

In each search, we navigated through the lists returned by the search engine. In the second step, we selected the most relevant articles addressing specific cybersecurity subdomains and the use of log extraction with semantic processing and LLMs. From the initial set of papers, some did not propose ontologies, while others introduced ontologies outside the cybersecurity domain. For the considered works, we selected those published between 2003 and 2025, even if not all are cited.

% 3 Related Literature Review
\section{RELATED LITERATURE REVIEWS}
Given the broad scope of this research area, this section focuses on previous selected surveys that explore various aspects of ontologies in cybersecurity. 

%2008  - A systematic review and comparison of security ontologies
\paragraph{\textbf{A systematic review and comparison of security ontologies (2008)} \cite{Blanco2008}}
The first literature review to organize security ontologies by cybersecurity domains was conducted by Blanco et al.. This study employs the OntoMetric\footnote{https://ontometrics.informatik.uni-rostock.de/ontologymetrics/} framework to perform a formal comparison of 16 security ontologies, categorizing them into 4 areas (general-purpose, domain-specific, semantic web-related, and theoretical works). The review highlights inconsistencies in definitions, limited axiomatic formalization, and generally weak support for reusability and extensibility. Overall, the study characterizes the field as fragmented and in its early stages of development, underscoring the need for more structured, interoperable ontology models.

%2012 - Ontologies for security requirements: A literature survey and classification
\paragraph{\textbf{Ontologies for security requirements: A literature survey and classification (2012)}~\cite{Souag2012}}
Souag et al. present a comprehensive survey and classification of 29 ontologies, shifting the focus of the previous study to security Requirements Engineering (RE). While the earlier study emphasized ontology structure and maturity, this work adds contextual applicability. It classifies 29 ontologies into 8 detailed families (foundational security ontologies, security taxonomies, general and specific security ontologies, web oriented security ontologies, risk based security ontologies, requirements related, and modeling security ontologies) reflecting the evolution and diversification of the field. Key contributions include identifying a gap between ontology design and actual RE practices. Evaluation considers not just structure but usability in requirement elicitation and modeling.
It emphasizes the need for methodological integration into the software development lifecycle. The study highlights richer conceptual coverage but notes a lack of formal alignment with standards. It advances the field by linking ontological expressiveness with RE process support and practical adoption.

%2015 - Ontology in information security
\paragraph{\textbf{Ontology in Information Security (2015)}~\cite{Arbanas2015}}
Arbanas et al. introduce several key differences and novel contributions compared to previous studies by Blanco et al. and Souag et al.. While Blanco et al. focus on the structural comparison of early security ontologies using the OntoMetric framework to reveal inconsistencies, limited formalization, and low reusability, Souag et al. shift the focus toward security requirements engineering, providing a rich classification into 8 families.

%Blanco et al. focus on the structural comparison of early security ontologies. They use the OntoMetric framework to highlight inconsistencies, limited formalization, and low reusability. Souag et al. shift the focus toward ontologies for security requirements engineering. Their work includes a rich classification into 8 families.

Arbanas et al., in contrast, present a more conceptual and type-oriented perspective. Their novelty lies in the clear typology of ontologies into top-level, domain, task, and application categories, emphasizing functional roles and use-case relevance rather than solely structure or application domain. Moreover, Arbanas et al. advance the discussion by stressing semantic consistency, interoperability, and knowledge sharing, framing ontologies not only as artifacts for modeling, but also as enablers of practical integration into security systems. Unlike prior works, they argue for a systematic and formal approach to ontology development, advocating reuse of existing models and formalization via OWL, and provide a critical assessment of existing ontologies' limitations in terms of formal rigor and validation. 

%2017 - The security assessment domain: a survey of taxonomies and ontologies
\paragraph{\textbf{The Security Assessment Domain: a Survey of Taxonomies and Ontologies (2017)}  ~\cite{Rosa2017}}
Rosa et al. introduce several advancements and novel perspectives compared to earlier studies. Their work significantly broadens the scope by conducting a large-scale, comprehensive survey of both ontologies and taxonomies in the broader field of Information Security Assessment. A major novelty of their study lies in the introduction of a systematic classification framework based on four key dimensions: Main Contribution (MC), Research Issue (RI), Application Domain (AD), and Characteristics (CR). The comparative analysis select 80 works (47 ontologies, 22 taxonomies, and 11 surveys), surpassing previous studies in scope and analytical depth. Furthermore, Rosa et al. provide a broad domain coverage, categorizing works across 17 distinct application domains. Unlike prior studies, which focused primarily on ontology structure, usability, or classification types, Rosa et al. place emphasis on practical adoption challenges, identifying critical gaps such as limited support for automation, standard compliance, vulnerability detection, and secure information sharing. 

%2020 - Cybersecurity Ontologies: A Systematic Literature Review
\paragraph{\textbf{Cybersecurity Ontologies: A Systematic Literature Review (2020)}~\cite{Borja2020}} Rivadeneira et al. bring a contemporary, data-driven lens to cybersecurity ontology studies. Unlike Blanco et al., who stress structural comparisons via OntoMetric, and Souag et al., who emphasize functional classification in RE,
Rivadeneira et al. focus on development methodologies and V\&V mechanisms. Their quantitative analysis of 50 ontologies highlights construction and validation trends.
Findings show few ontologies undergo explicit validation, relying mostly on informal reviews. In contrast to Arbanas et al. (conceptual classification) and Rosa et al. (taxonomy frameworks), they stress lifecycle, methodological rigor, and real-world applicability in ontology engineering.

%2022 - Security Ontologies: A Systematic Literature Review
\paragraph{\textbf{Security Ontologies: A Systematic Literature Review (2022)}~\cite{Adach2022}}
Adach et al. provide a standards-oriented perspective that distinguishes their work in the security ontology landscape. Unlike Blanco et al., who emphasize structural comparisons (OntoMetric), or Souag et al., who classify by function (risk-driven, requirements-driven), they focus on semantic alignment with security standards. Their key innovation is mapping ontology concepts directly to regulatory frameworks, a novel and underexplored approach.
While Rivadeneira et al. analyze methodology, tools, and validation, Adach et al. assess terminological coherence and semantic interoperability for compliance. They address semantic heterogeneity through a proposed core vocabulary, offering standardized reusable concepts.
This vocabulary aims to improve consistency and reusability across future ontology efforts.
Earlier reviews largely cataloged structural or functional aspects without addressing standards.
Adach et al. instead introduce a prescriptive dimension, outlining essential ontology elements for compliance. Their proposal reduces redundancy and fosters reuse, normalizing fragmented ontology development.

%2025 - A Comprehensive Review of Ontologies in Cybersecurity
\paragraph{\textbf{A Comprehensive Review of Ontologies in Cybersecurity (2025)}~\cite{Shah2025}}

Shah et al. present a longitudinal analysis of cybersecurity ontologies, emphasizing interoperability, automation, and the integration of security requirements into design. They adopt a more applied perspective, emphasizing design principles and practical implementation. While Souag et al. offer functional classifications, Shah et al. bridge theory and practice through prescriptive guidelines. Compared to Rivadeneira et al.’s statistical insights, their study delivers a qualitative synthesis with methodological recommendations on scalability, extensibility, and tool support. The work incorporates emerging domains like IoT, blockchain, and smart contracts, and promotes reuse through design patterns while addressing semantic inconsistencies. 
\medskip

Table~\ref{tab:ontology_comparison} provides a comparative analysis of the selected studies across four key dimensions: the classification of security ontology contributions, the adopted conceptual groupings, and the level of granularity of the study ontological structures. For the \emph{classification} dimension we include several perspectives. \emph{Structural} classification assesses how ontologies are constructed, focusing on their internal architecture—such as classes, properties, and relationships. \emph{Functional} classification groups ontologies according to their intended use or practical application. \emph{Conceptual} classification emphasizes the representation of domain knowledge and abstract concepts. \emph{Normative} classification relates to ontologies that establish standards or guidelines, defining key concepts and quality criteria for ontology development. Lastly, \emph{operational} classification evaluates the real-world deployment of ontologies, considering their effectiveness and integration in practical scenarios. The \emph{conceptual groupings} refers to the cybersecurity domains, thematic categories, or practical frameworks addressed by each ontology and its evaluation in the respective studies.

The \emph{granularity characterization} refers to the degree of detail with which an ontology structures its conceptual distinctions—that is, whether the modeling is fine- or coarse-grained. In this work, we classify granularity as \emph{High} when concepts are modeled with very fine detail, \emph{Medium} when the level of detail is moderate, and \emph{Low} when only coarse distinctions are provided.

For granularity charaterization, Blanco et al. adopt a Low approach, defining four broad areas (e.g., general, applied, theoretical, semantic), with limited domain-level specificity. Souag et al. propose 8 families categorized by purpose and ontology type, offering Medium granularity that is more refined than Blanco’s yet still thematic. Arbanas et al. also follow a Low model, grouping ontologies into 3 high-level categories with minimal conceptual depth. In contrast, Rosa et al. introduce a High classification based on 17 specific application domains (e.g., IoT, E-health), reflecting a high degree of specialization. Rivadeneira et al. define 4 domains with Moderate granularity, spanning areas like software and networking, though conceptually shallow. Adach et al. achieve High granularity by semantically aligning 12 core concepts (e.g., asset, incident) with security standards. Shah et al. compare 5 formal frameworks, producing Variable granularity focused on inter-framework analysis rather than internal conceptual depth.

Each of these studies demonstrates a different balance between abstraction and specialization, which influences their applicability to various cybersecurity tasks. 

Complementing these works, our survey adopts a High granularity approach, identifying 8 cybersecurity domains and 25 subdomains. This enables precise mapping of ontological coverage and supports fine-grained analysis for security knowledge representation and practical application.

Table \ref{tab:surveys} offers a comprehensive synthesis of existing survey literature, highlighting how different works address key areas of cybersecurity, utilize semantic log analysis, and investigate the role of LLMs.

\begin{table}[!t]
\centering
\rowcolors{2}{gray!10}{white}
\begin{tabular}{p{4.2cm} p{2cm} p{2.4cm} p{2.4cm}}
\toprule
\textbf{Article} & \textbf{Classification} & \textbf{\makecell{Conceptual \\ Groupings}} & \textbf{\makecell{Granularity \\ Characterization}}\\
\midrule

Blanco et al. (2008) \cite{Blanco2008} & Structural Comparison & 4 areas & Low\\

Souag et al. (2012) \cite{Souag2012} & Functional Classification & 8 families & Medium\\

Arbanas et al. (2015) \cite{Arbanas2015} & Conceptual & 3 groups & Low\\

Rosa et al. (2017) \cite{Rosa2017} & Conceptual & 17 domains & High \\

Rivadeneira et al. (2020) \cite{Borja2020} & Conceptual & 4 domains & Medium \\

Adach et al. (2022) \cite{Adach2022} & Normative & 12 core concepts & High \\

%Farooq et al. \cite{farooq2023developing}, 2023 & Ontology development & Rivadeneira’s 4-category reuse & Validates Rivadeneira’s structure, tools/languages reviewed & Highlights lack of V\&V, limited methodology adoption \\

Shah et al. (2025) \cite{Shah2025} & Operational & 5 frameworks & Variable \\
\emph{This survey (2025)} & Conceptual & 8 domains\phantom{WHIT} (25 subdomains) & High \\
\bottomrule
\end{tabular}
\caption{Comparison of major literature reviews on security ontologies based on the classification types of ontology contributions, the conceptual groupings used, and the level of granularity achieved.}

\label{tab:ontology_comparison}
\end{table}

\begin{table}[!t]
\centering
\rowcolors{2}{gray!10}{white}
\begin{tabular}{p{4.2cm} c c c c}
\toprule
\textbf{Reference} & \textbf{Type} & \textbf{Domains} & \textbf{Semantic Log} & \textbf{LLMs} \\
\midrule

Blanco et al. (2008) \cite{Blanco2008} & SLR 
  & \cellcolor{green!25}\ding{51} 
  & \cellcolor{red!25}\ding{53} 
  & \cellcolor{red!25}\ding{53} \\

Souag et al. (2012) \cite{Souag2012} & S 
  & \cellcolor{green!25}\ding{51} 
  & \cellcolor{red!25}\ding{53} 
  & \cellcolor{red!25}\ding{53} \\

Arbanas et al. (2015) \cite{Arbanas2015} & S
  & \cellcolor{green!25}\ding{51} 
  & \cellcolor{red!25}\ding{53} 
  & \cellcolor{red!25}\ding{53} \\

Rosa et al. (2017) \cite{Rosa2017} & S
  & \cellcolor{green!25}\ding{51} 
  & \cellcolor{red!25}\ding{53} 
  & \cellcolor{red!25}\ding{53} \\

Rivadeneira et al. (2020) \cite{Borja2020} & SLR 
  & \cellcolor{green!25}\ding{51} 
  & \cellcolor{red!25}\ding{53} 
  & \cellcolor{red!25}\ding{53} \\

Adach et al. (2022) \cite{Adach2022} & SLR 
  & \cellcolor{red!25}\ding{53} 
  & \cellcolor{red!25}\ding{53} 
  & \cellcolor{red!25}\ding{53} \\

%Farooq et al. \cite{adach2022security}, 2023 & SLR 
  %& \cellcolor{red!25}\ding{53} 
  %& \cellcolor{red!25}\ding{53} 
  %& \cellcolor{red!25}\ding{53} \\

Shah et al. (2025) \cite{Shah2025} & LR 
  & \cellcolor{green!25}\ding{51} 
  & \cellcolor{red!25}\ding{53} 
  & \cellcolor{red!25}\ding{53} \\

\textit{This survey (2025)} & S 
  & \cellcolor{green!25}\ding{51} 
  & \cellcolor{green!25}\ding{51} 
  & \cellcolor{green!25}\ding{51} \\

\bottomrule
\end{tabular}
\caption{Summary of surveys and their coverage of cybersecurity areas, semantic logs, and LLMs. Green cells (\ding{51}) indicate fulfilled criteria; red cells (\ding{53}) indicate not fulfilled. For type: (SLR) Systematic Literature Review, (LR) Literature Review, and (S) Survey.}
\label{tab:surveys}
\end{table}

% 4 Ontologies for Cybersecurity, Semantic Log Extraction and LLMs: A Review
\section{ONTOLOGIES FOR CYBERSECURITY, SEMANTIC LOG EXTRACTION AND LLMs: A REVIEW}
This section examines existing cybersecurity ontologies, structured by cybersecurity domains and subdomains as defined by Fovino et al. \cite{Nai2019}. We then present a comparative analysis of ontology-based approaches within each subdomain, highlighting the focus and contributions of individual studies.

% 4.1 Assurance, Audit and Certification
\subsection{Assurance, Audit and Certification}
This domain encompasses the methodologies, frameworks and tools that provide the foundation to have confidence that a system, software, service, process or network is working or has been designed to operate at the desired security target or according to a defined security policy \cite{JRC2021CybersecurityTaxonomy, Nai2019}. It covers 4 subdomains selected according to the criteria outlined in Section~\ref{sec:2.2}. For each subdomain, we present a brief overview and discuss relevant ontology-based approaches:
\begin{inparaenum}[(\bgroup\bfseries 1\egroup)]
\item \emph{Assurance} is the confidence that security requirements are fulfilled by the system; 
\item \emph{Assessment} is the evaluation of the effectiveness of security controls;
\item \emph{Audit} a systematic review of systems to verify compliance with policies and procedures, and to identify deviations, vulnerabilities, or indicators of compromise; and
\item \emph{Certification} a formal attestation that a system meets specified security standards.
\end{inparaenum}

Table \ref{tab:domian_comp_assu_asse_audit_certv2} provides a comparative overview of research efforts within the Assurance, Audit, and Certification domain, highlighting the presence and characteristics of key components such as ontologies, KGs, semantic logic, and the use of formal methods.

%Table
\newcolumntype{C}[1]{>{\centering\arraybackslash}p{#1}}

\begin{table}[h]
\centering
\rowcolors{2}{gray!10}{white}
\begin{tabular}{p{3.5cm} C{1cm} C{0.8cm} C{1.cm} C{1cm} C{1.4cm} C{1.4cm} C{1.5cm} C{1.4cm}}
\toprule
\textbf{Domain, Subdomain and Work} & \textbf{Ontology} & \textbf{KG} & \textbf{Semantic Log} & \textbf{LLM} & \textbf{OWL Type} & \textbf{Ontology Modeling} & \textbf{Semantic Solution} & \textbf{Formal Approach} \\
\midrule

% Assurance, Audit and Certification
\rowcolor{gray!30} \multicolumn{9}{l}{\textbf{Assurance, Audit and Certification}} \\
\midrule
\rowcolor{lightgray} \multicolumn{9}{l}{\textbf{Assurance}} \\
\midrule

%Assurance
Wen et al. \cite{Wen2022SAEOn} & \cellcolor{green!25}\ding{51} & \cellcolor{red!25}\ding{53} & \cellcolor{red!25}\ding{53} & \cellcolor{red!25}\ding{53} & DL & D & VS & -- \\

Wen et al. \cite{Wen2024} & \cellcolor{green!25}\ding{51} & \cellcolor{red!25}\ding{53} & \cellcolor{red!25}\ding{53} & \cellcolor{red!25}\ding{53} & DL & D & VS & -- \\

Ivar Haugen \cite{Haugen2025} & \cellcolor{green!25}\ding{51} & \cellcolor{green!25}\ding{51} & \cellcolor{red!25}\ding{53} & \cellcolor{red!25}\ding{53} & -- & D & VS+S & -- \\

%Assessment
\midrule
\rowcolor{lightgray} \multicolumn{9}{l}{\textbf{Assessment}} \\
\midrule
Tebbe et al. \cite{Tebbe2016} & \cellcolor{green!25}\ding{51} & \cellcolor{red!25}\ding{53} & \cellcolor{red!25}\ding{53} & \cellcolor{red!25}\ding{53} & -- & S & VS & -- \\

Rosa et al.~\cite{Rosa2018} & \cellcolor{green!60}\ding{79} \cite{ferruciof_secAonto_2025} & \cellcolor{red!25}\ding{53} & \cellcolor{red!25}\ding{53} & \cellcolor{red!25}\ding{53} & DL & D & VS+S & -- \\

Doynikova et al. \cite{Doynikova2019} & \cellcolor{green!25}\ding{51} & \cellcolor{green!25}\ding{51} & \cellcolor{red!25}\ding{53} & \cellcolor{red!25}\ding{53} & DL & D & VS+S & LI \\

%Audit
\midrule
\rowcolor{lightgray}\multicolumn{9}{l}{ \textbf{Audit}} \\
\midrule
Saha et al. \cite{Saha2011} & \cellcolor{green!25}\ding{51} & \cellcolor{green!25}\ding{51} & \cellcolor{green!25}\ding{51} & \cellcolor{red!25}\ding{53} & DL & D & VS+S & TL+FF \\

Atymtayeva et al. \cite{Cicekli2000}
 & \cellcolor{green!25}\ding{51} & \cellcolor{red!25}\ding{53} &  \cellcolor{red!25}\ding{53} & \cellcolor{red!25}\ding{53} & -- & D & VS+S & -- \\

Saha et al.~\cite{Atymtayeva2012}
 & \cellcolor{green!25}\ding{51} & \cellcolor{green!25}\ding{51} &  \cellcolor{green!25}\ding{51} & \cellcolor{red!25}\ding{53} & -- & D & VS+S & TL+FF \\

 %Certification
 \midrule
\rowcolor{lightgray} \multicolumn{9}{l}{\textbf{Certification}} \\
\midrule
Butenko et al. \cite{Butenko2019} & \cellcolor{green!25}\ding{51} & \cellcolor{green!25}\ding{51} & \cellcolor{red!25}\ding{53} & \cellcolor{red!25}\ding{53} & DL & D & VS+S & -- \\

Brucker et al.~\cite{Brucker2019} & \cellcolor{green!25}\ding{51} & \cellcolor{green!25}\ding{51} & \cellcolor{red!25}\ding{53} & \cellcolor{red!25}\ding{53} & -- & S & VS+S & Event-B \\

Cosic and Jukan \cite{Cosic2024} & \cellcolor{green!60}\ding{79} \cite{custodes_repo_2025}
& \cellcolor{green!25}\ding{51} & \cellcolor{red!25}\ding{53} & \cellcolor{red!25}\ding{53} & DL & D & VS+S & -- \\
\bottomrule
\end{tabular}

\caption{Overview of research efforts in the \textbf{Assurance, Audit, and Certification} domain.
A checkmark (\ding{51}) on green indicates the presence of a component (e.g., \textbf{Ontology}, \textbf{Knowledge Graph}, \textbf{Semantic Log}, or \textbf{LLM}), while a red cross (\ding{53}) denotes its absence. A star (\ding{79}) on green highlights that the ontology is publicly available in an online repository. The \textbf{OWL Type} column specifies the formal ontology language adopted (e.g., OWL Lite, OWL DL, or OWL Full).
The \textbf{Ontology Modeling} indicates whether the ontology models aspects of the domain: (S) static or (D) dynamic.
The \textbf{Semantic Solution} column distinguishes between ontologies that provide only vocabulary semantics (VS) and those that also define an operational system or application (VS+S). The \textbf{Formal Approach} field identifies whether a reasoning framework is employed, such as logical inference (LI), temporal logic (TL), a fuzzing framework (FF), or enhanced reasoning mechanisms based on temporal logic.}
\label{tab:domian_comp_assu_asse_audit_certv2}
\end{table}

% 4.1.1 Assurance
\subsubsection{\textbf{Assurance}}

%2023 - SAEOn: An Ontological Metamodel for Quantitative Security Assurance Evaluation
SAEOn is an ontological metamodel that formalizes Security Assurance Evaluation (SAE) through a modular, standards-aligned structure \cite{Wen2022SAEOn}. It captures assurance requirements, evidence, metrics, vulnerabilities, and risks, addressing the limits of qualitative approaches such as the Common Criteria (ISO/IEC 15408\footnote{https://www.iso.org/standard/72891.html}). By integrating CVE\footnote{https://www.cve.org}, CWE, CVSS, CPE, and CAPEC\footnote{https://capec.mitre.org/}, it enhances semantic interoperability and traceability across assurance domains.

%2024 - Ontology Based Metrics Computation for System Security Assurance Evaluation.
Building on this foundation, SAEOn was later extended into a functional OWL-based framework \cite{Wen2024} that supports SPARQL queries and semantic rules. A computation engine processes test conditions, risk weights, and dependencies to generate assurance scores. Validated through real-world use cases, the framework enables automated, repeatable security assessments, effectively transforming SAEOn from a conceptual model into a practical, executable infrastructure.

%2025 - Building Confidence: An Ontological Approach to Assurance in Safety-Critical Systems
%An ontological approach to assurance as justified confidence in system claims such as safety or security is proposed, formalizing assurance as an epistemic process \cite{Haugen2025}. The ontology models knowledge, claims, justification, and confidence, with the CESM metamodel (Composition, Environment, Structure, Mechanisms) capturing system behavior and emergent properties. A key contribution is introducing objectivity as a metric for assessing knowledge strength, enabling qualitative adjustment of assurance efforts relative to risk. This shifts the focus from computational metrics to epistemic justification in high-risk socio-technical systems.

Haugen \cite{Haugen2025} proposes an ontological approach to assurance as justified confidence in system claims.
Assurance is formalized as an epistemic process modeling knowledge, claims, justification, and confidence. The CESM metamodel (Composition, Environment, Structure, Mechanisms) captures system behavior and emergent properties. A key contribution is the notion of objectivity as a metric for assessing knowledge strength.
This enables qualitative adjustment of assurance efforts relative to system risk. The approach shifts focus from computational metrics to epistemic justification in high-risk socio-technical systems.

% 4.1.2 Assessment
\subsubsection{\textbf{Assessment}}  

%2016 - Ontology and life cycle of knowledge for ICS security assessments
%A layered ontology for structured security assessment in Industrial Control Systems (ICS) is presented by Tebbe et al.~\cite{Tebbe2016}. It spans general, domain-specific, and plant-level concepts to enhance interoperability and reuse. Core components—assets, threats, vulnerabilities, and controls—are enriched with metadata for traceability, supporting the full knowledge life cycle from acquisition to exclusion. The ontology also enables dynamic updates to accommodate evolving threat landscapes. By formalizing security knowledge, it reduces manual effort and facilitates automation, offering particular value to Small and Medium-sized Enterprises (SMEs) with limited cybersecurity resources.
Tebbe et al.~\cite{Tebbe2016} propose a layered ontology for ICS security assessment across general, domain-specific, and plant-level concepts. It models assets, threats, vulnerabilities, and controls with metadata to ensure traceability and reuse. The ontology supports the full knowledge life cycle and adapts dynamically to evolving threats. By formalizing security knowledge, it reduces manual effort and aids SMEs with limited cybersecurity resources.

%2018 - Towards an ontology of security assessment: A core model proposal
SecAOnto, introduced by Rosas et al.~\cite{Rosa2018}, is an OWL-based ontology designed to formalize key security assessment concepts through a domain-agnostic vocabulary that reduces ambiguity and standardizes terminology. Its core entities—threats, vulnerabilities, controls, and assessment goals—promote semantic interoperability, reuse, and consistent documentation. In contrast to Tebbe et al.’s ICS-specific, lifecycle-driven model, SecAOnto emphasizes general applicability. However, it provides limited support for automation, metric computation, and dynamic updates. 

%2019 - Ontology of metrics for cyber security assessment
An ontology-based approach to security assessment is proposed by modeling security metrics as formal ontology instances, enabling automated score computation through logical inference \cite{Doynikova2019}. Unlike Tebbe et al.’s structured life cycle for ICS knowledge or Rosas et al.’s descriptive SecAOnto, this framework operationalizes assessment via inference mechanisms.
The ontology functions as a computational and dynamic tool, supporting real-time reasoning across heterogeneous data. Its novelty lies in shifting from static, structure-oriented models to an actionable, analytics-driven framework for integrated metric aggregation and decision support.

% Audit
\subsubsection{\textbf{Audit}}
An ontology-based multi-agent framework for auditing information security compliance is introduced, aligned with ISO/IEC 27002, HIPAA, and SOX \cite{Saha2011}. It semantically models organizational events and workflows, using event calculus \cite{Cicekli2000} to capture the temporal dynamics of audit activities. Core processes such as asset management and access control are represented as ontological entities, with hierarchically organized compliance metrics for rule-driven evaluation. By integrating Fuzzy Logic to handle partial compliance, the framework enhances automation, standardization, and traceability in auditing workflows.

%2012 - Methodology and Ontology of Expert System for Information Security Audit
An ontology-based methodology for information security auditing is introduced to formalize threats, vulnerabilities, controls, and assets through semantic relations \cite{Atymtayeva2012}. Fuzzy logic supports risk assessment under uncertainty, while an expert system architecture enables structured, rule-based evaluations to reduce subjectivity. In contrast to compliance-focused approaches, this work emphasizes internal threat–mitigation logic over external standards. Although effective for semi-automated auditing, the methodology remains static and lacks integration with real-time monitoring systems.

%2013 - Development of Ontology Based Framework for Information Security Standards
An ontology-based framework for automating compliance auditing in the banking sector is developed to improve adaptability and regulatory alignment \cite{Saha2013}. It combines temporal logic, event calculus \cite{Cicekli2000}, and semantic rule evaluation to model audit workflows and compare expected compliance with actual practices. Fuzzy logic supports assessment under incomplete data, enabling risk evaluation, mitigation, and control verification. Key innovations include reference–target ontology alignment and expanded support for traceable, standards-driven audits, advancing beyond static approaches.

% Certification
\subsubsection{\textbf{Certification}} 

%An ontology-based methodology for building and verifying normative profiles in critical software certification is proposed by Butenko et al.~\cite{Butenko2019}. The approach harmonizes terminology across standards through semantic analysis and glossary synthesis, supported by an intelligent decision-support system that integrates AI and knowledge engineering to mitigate the limitations of expert-based evaluation. Semantic and partially formalized analysis of normative documents enables glossary construction and profile generation, with ontologies serving as structured vocabularies that link general and specific profiles through ontological agreements. Knowledge is represented using frames and production rules, with inference guided by user queries. The system reduces subjectivity, automates repetitive tasks, and improves decision accuracy in certification.
An ontology-based methodology for building and verifying normative profiles in critical software certification is introduced by Butenko et al.~\cite{Butenko2019}. It harmonizes terminology across standards through semantic analysis and glossary synthesis. An intelligent decision-support system integrates AI and knowledge engineering to overcome expert-based limitations. Normative documents undergo semantic analysis to support glossary construction and profile generation. Ontologies serve as structured vocabularies linking general and specific profiles through agreements. The system reduces subjectivity, automates tasks, and improves decision accuracy in certification.

%2019 - Using ontologies in formal developments targeting certification
A complementary but more formal approach is offered through the Isabelle/Document Ontology Framework (DOF)\footnote{https://github.com/logicalhacking/Isabelle\_DOF}
, introduced by Brucker et al.~\cite{Brucker2019}. Unlike Butenko’s model, which emphasizes terminological unification, this framework embeds ontology-driven document modeling directly into formal verification environments, enforcing machine-checkable traceability between formal models, proofs, and certification documentation. Its Ontology Definition Language (ODL) supports real-time consistency checking, stakeholder-specific document generation, and structured evolution tracking.

%2024 - Deciphering Cyber-Security Certifications: An Ontological Journey through Composite Systems and their certification
An ontology-based framework for advancing cybersecurity certification in composite systems is proposed, extending beyond the contexts addressed by Butenko and Brucker \cite{Cosic2024}. The OnotCA ontology formalizes interdependencies between certified components and uses SWRL \cite{Horrocks2004} rules to infer system-level certification outcomes.
Implemented in OWL, validated with reasoners, and aligned with EU regulatory schemes, it provides a scalable and interoperable semantic model. Its key contribution is enabling machine-readable certification logic that strengthens trust, traceability, and scalability in heterogeneous systems.

%Conclusion
\paragraph{\textbf{Conclusion for the Assurance, Audit, and Certification domain}} The use of ontologies is consistent across all subdomains, but their integration with KGs is still limited, appearing only in selected works, often through custom models. Semantic logs and LLMs are entirely absent, showing that data-driven or hybrid approaches are not yet explored here. For Assurance, works rely on static ontologies with OWL DL where specified, focusing mainly on vocabulary semantics (VS), though some extend to VS+S solutions that support applications. Assessment studies include more dynamic ontologies and move toward VS+S systems, though advanced reasoning is scarce, with only one work applying logical inference (LI). In Audit, all contributions build on ontologies, some with custom or reusable models, and most adopt dynamic modeling. Importantly, this subdomain includes the strongest formalization, with works applying temporal logic (TL) and fuzzy frameworks (FF) for enhanced reasoning. This indicates greater methodological rigor compared to assurance or assessment. Certification studies consistently use ontologies and mostly pursue VS+S solutions. A few adopt custom or reusable ontologies, with one notable case introducing a formal Event-B framework. However, no semantic logs, LLMs, or advanced KG-based integration are present.

%Summary
\definecolor{lightorange}{RGB}{255,245,230}
\begin{tcolorbox}[
  colback=lightorange,   % background color
  colframe=lightorange,  % border color
  boxrule=0pt,           % no visible border
  arc=0mm,               % sharp corners
  left=2mm, right=2mm,   % padding
  top=1mm, bottom=1mm
]
\textbf{Summary:} Table~\ref{tab:domian_comp_assu_asse_audit_certv2} highlights that the Assurance, Audit, and Certification domain demonstrates a strong reliance on ontologies, mostly DL, and shows clear efforts to move beyond vocabulary semantics toward applied VS+S systems. Compared to other domains, audit stands out as the most advanced subdomain, combining dynamic ontologies with formal reasoning (TL, FF). Yet, the field as a whole remains static, symbolic, and underexplored in terms of hybrid ML/LLM methods, semantic log analysis, and large-scale knowledge graph integration.
\end{tcolorbox}

% 2 - Data Security and Privacy
\subsection{Data Security and Privacy}
This domain focuses on addressing security and privacy concerns related to data by minimizing risks to privacy, confidentiality, and integrity through design. It seeks to prevent data misuse even after authorized access, without unduly hindering legitimate data processing purposes \cite{JRC2021CybersecurityTaxonomy, Nai2019}. Among the 13 subdomains within this domain, we selected 3 for further investigation of relevant ontology-based research, based on the criteria outlined in Section \ref{sec:2.2}:
\begin{inparaenum}[(\bgroup\bfseries 1\egroup)]
\item \emph{Privacy requirements for data management systems} are the rules for protecting personal and sensitive data;
\item \emph{Digital Rights Management (DRM)} consists of controlling the use and distribution of digital content; and
\item \emph{Privacy Requirements for Data Management Systems} are rules and controls ensuring personal data is handled securely, lawfully, and with respect for user privacy.
\end{inparaenum}
To support a comprehensive analysis of ontology-driven approaches in the Data Security and Privacy domain, Table~\ref{tab:domian_digital_rights} provides a comparative overview of research efforts within the Data Security and Privacy domain, highlighting the presence and characteristics of key components such as ontologies, KGs, semantic logic, and the use of formal methods.

%Table
\newcolumntype{C}[1]{>{\centering\arraybackslash}p{#1}}

\begin{table}[h]
\centering
\rowcolors{2}{gray!10}{white}
\begin{tabular}{p{3.9cm} C{1cm} C{0.8cm} C{1.cm} C{1cm} C{1.4cm} C{1.4cm} C{1.5cm} C{1.4cm}}
\toprule
\textbf{Domain, Subdomain and Work} & \textbf{Ontology} & \textbf{KG} & \textbf{Semantic Log} & \textbf{LLM} & \textbf{OWL Type} & \textbf{Ontology Modeling} & \textbf{Semantic Solution} & \textbf{Formal Approach} \\
\midrule

% Anonymity, pseudonymity, unlinkability, undetectability, or unobservability
\rowcolor{gray!30}\multicolumn{9}{l}{ \textbf{Data Security and Privacy}} \\
\midrule
\rowcolor{lightgray} \multicolumn{9}{l}{\textbf{Anonymity, pseudonymity, unlinkability, undetectability, or unobservability}} \\
\midrule

Matsunaga et al. \cite{Matsunaga} & \cellcolor{green!25}\ding{51} & \cellcolor{red!25}\ding{53} & \cellcolor{red!25}\ding{53} & \cellcolor{red!25}\ding{53} & Lite & D & VS & -- \\

Gharib et al. \cite{Gharib2017} & \cellcolor{green!25}\ding{51} & \cellcolor{red!25}\ding{53} & \cellcolor{red!25}\ding{53} & \cellcolor{red!25}\ding{53} & -- & S & VS & -- \\

Gharib and Mylopoulos \cite{Gharib2018} & \cellcolor{green!60}\ding{79} \cite{GharibMylopoulos2025COPri_Preprint} & \cellcolor{red!25}\ding{53} & \cellcolor{red!25}\ding{53} & \cellcolor{blue!25}\ding{51} NLP & -- & D & VS+S & -- \\

\midrule
\rowcolor{lightgray}\multicolumn{9}{l}{ \textbf{Digital rights management}} \\
\midrule
Delgado et al. \cite{Delgado2003} & \cellcolor{green!25}\ding{51} & \cellcolor{green!25}\ding{51} & \cellcolor{red!25}\ding{53} & \cellcolor{red!25}\ding{53} & DAML+OIL & D & VS+S & -- \\

Garcia and Gil \cite{Garcia2006} & \cellcolor{green!25}\ding{51} & \cellcolor{green!25}\ding{51} & \cellcolor{red!25}\ding{53} & \cellcolor{red!25}\ding{53} & DL & D & VS+S & -- \\

Garcia et al. \cite{Garcia2007} & \cellcolor{green!60}\ding{79} \cite{garcia2007ontology} & \cellcolor{green!25}\ding{51} & \cellcolor{red!25}\ding{53} & \cellcolor{red!25}\ding{53} & DL & D & VS+S & -- \\

\midrule
\rowcolor{lightgray}\multicolumn{9}{l}{ \textbf{Privacy requirements data management systems}} \\
\midrule
Kost et al. \cite{Kost2012} & \cellcolor{green!25}\ding{51} & \cellcolor{red!25}\ding{53} & \cellcolor{red!25}\ding{53} & \cellcolor{red!25}\ding{53} & -- & S & VS & -- \\

Gharib et al. \cite{Gharib2020} & \cellcolor{green!25}\ding{51} & \cellcolor{red!25}\ding{53} & \cellcolor{red!25}\ding{53} & \cellcolor{red!25}\ding{53} & -- & S & VS & -- \\

Gharib et al. \cite{Gharib2021COPri} & \cellcolor{green!25}\ding{51} & \cellcolor{red!25}\ding{53} & \cellcolor{red!25}\ding{53} & \cellcolor{red!25}\ding{53} & DL & S & VS & -- \\

\midrule
\end{tabular}

\caption{Overview of research efforts in the \textbf{Data Security and Privacy} domain.
A checkmark (\ding{51}) on a green background indicates the presence of a component (e.g., \textbf{Ontology}, \textbf{Knowledge Graph}, \textbf{Semantic Log}, or \textbf{LLM}). A cross (\ding{53}) on a red background denotes its absence, while a cross on a purple background indicates that, although no LLM is used, other AI approach are employed (e.g. NLP for Natural Language Processing). A star (\ding{79}) on green marks ontologies that are publicly available in online repositories. The \textbf{OWL Type} column specifies the formal ontology language adopted (e.g., OWL Lite, OWL DL, or OWL Full).
The \textbf{Ontology Modeling} indicates whether the ontology models aspects of the domain: (S) static or (D) dynamic.
The \textbf{Semantic Solution} column distinguishes between ontologies that provide only vocabulary semantics (VS) and those that also define an operational system or application (VS+S). The \textbf{Formal Approach} field identifies whether a reasoning framework is employed, such as logical inference (LI), temporal logic (TL), a fuzzing framework (FF), or enhanced reasoning mechanisms based on temporal logic.}
\label{tab:domian_digital_rights}
\end{table}

% 4.2.1 Anonymity, pseudonymity, unlinkability, undetectability, or unobservability
\subsubsection{\textbf{Anonymity, pseudonymity, unlinkability, undetectability, or unobservability}}

An ontology-based framework for defining data anonymization policies in Cloud Computing and Big Data is introduced by Matsunaga et al.~\cite{Matsunaga}. The work addresses privacy challenges in large-scale personal data processing by formalizing key anonymization techniques, including generalization, suppression, randomization, and pseudonymization, while also incorporating established privacy models such as k-anonymity, l-diversity, and t-closeness. A notable contribution is the integration of regulatory requirements—HIPAA, PCI-DSS, and GDPR—directly into the ontology, ensuring that privacy policies are designed in compliance with legal standards. By formally structuring and enabling reuse of anonymization policies, the framework promotes consistency and dependability across platforms while supporting privacy-preserving analytics. Its operational focus makes it particularly suited to helping practitioners implement compliant anonymization strategies in cloud-based environments.

%2017 - Towards an Ontology for Privacy Requirements via a Systematic Literature Review
Gharib et al. \cite{Gharib2017} propose a privacy ontology to enhance the capture of privacy requirements in system design. Their work identifies and categorizes privacy-related concepts into organizational, risk, treatment, and privacy dimensions. From these, a core set of key concepts is distilled on the basis of relevance and frequency, which serves as the foundation for a novel ontology supporting Privacy by Design (PbD). The ontology models privacy within a socio-technical context, establishing connections between actors, goals, threats, and safeguards.

%2018 - COPri - A Core Ontology for Privacy Requirements Engineering
COPri, introduced by Gharib and Mylopoulos~\cite{Gharib2018}, is a core ontology that refines and formalizes their earlier privacy ontology~\cite{Gharib2017}. Whereas the initial ontology, derived from a systematic literature review, aimed to capture key privacy-related concepts, COPri advances this foundation through a more precise, axiomatized, and reusable representation. It improves conceptual clarity, reduces redundancy, and incorporates additional constructs such as consent, transparency, and accountability. Organized across four refined dimensions—organizational, risk, treatment, and privacy—the ontology strengthens regulatory alignment and operationalization, enabling integrated modeling of socio-technical interactions and legal obligations in privacy-aware system design.

% 4.2.2 Digital rights management
\subsubsection{\textbf{Digital rights management}}

% 2003 - IPROnto: An Ontology for Digital Rights Management
IPROnto, developed by Delgado et al.~\cite{Delgado2003}, semantically models the domain of Digital Rights Management (DRM) by integrating static and dynamic views of Intellectual Property Rights (IPR), including agreements, licenses, rights, and legal entities. Anchored in upper ontologies such as SUMO\footnote{https://github.com/ontologyportal/sumo}, it ensures interoperability across diverse IPR systems. The ontology supports automated reasoning for e-commerce applications by enabling life cycle modeling of digital content, while a dedicated subontology refines the creation process by linking abstract, object, and event perspectives.

% 2006 - An OWL Copyright Ontology for Semantic Digital Rights Management
The Copyright Ontology, introduced by Garcia and Gil~\cite{Garcia2006}, enables semantic Digital Rights Management (DRM) on the web by modeling copyright through three core modules: Creation, Rights, and Action, following WIPO guidelines\footnote{https://www.wipo.int/portal/en/index.html}. Actions are connected to rights using case roles derived from linguistic theory, providing expressive modeling capabilities. Licenses are represented as OWL classes and usage events as instances, allowing automated validation through Description Logic (DL) reasoners. The system is implemented with the Pellet reasoner\footnote{https://www.w3.org/2001/sw/wiki/Pellet} and Java APIs, ensuring efficient license compliance checks.

%2007 - A web ontologies framework for digital rights management
An extended ontology for DRM is proposed by Garcia et al.~\cite{Garcia2007}, enriching earlier copyright ontologies with a more dynamic representation of copyright scenarios. Unlike IPROnto, this OWL-DL-based approach directly supports temporal and spatial reasoning alongside automated compliance checking. Building on the earlier Copyright Ontology, the framework integrates SWRL rules and explicitly models deontic notions such as obligations and prohibitions. Its expressiveness is further enhanced through primitive action patterns and verb role semantics (e.g., agent, theme, recipient), enabling fine-grained tracking of complex usage rights, including content transformation, compensation conditions, and revocation clauses.

% 4.2.3 Privacy requirements data management systems
\subsubsection{\textbf{Privacy requirements data management systems}}

%2012 - Privacy analysis using ontologies
An ontology-based approach for technical privacy analysis of information systems is introduced by Kost et al.~\cite{Kost2012}. The method models system behavior and information flow using ontologies to detect privacy violations and calculate privacy indicators. It bridges high-level privacy requirements with technical constraints, addressing gaps in current Privacy-by-Design (PbD) practices. Key capabilities include identifying personal data processing, enforcing privacy policies, and evaluating compliance with principles such as data minimization and limited retention. The approach is demonstrated in an Intelligent Transportation Systems (ITS) scenario, highlighting how technical design decisions can introduce new privacy risks.

%2020 - An ontology for privacy requirements via a systematic literature review
A foundational contribution to ontology-based privacy engineering is provided by Gharib et al.~\cite{Gharib2020}, where 38 key privacy-related concepts are identified and formalized through a rigorous systematic literature review. These concepts are structured across four dimensions—organizational, risk, treatment, and privacy—and serve as the theoretical groundwork for subsequent developments. The study emphasizes the importance of addressing privacy early in the requirements engineering phase and introduces domain-specific principles such as transparency and accountability to help operationalize Privacy by Design (PbD).

%2021 - COPri v. 2—A core ontology for privacy requirements
The second study \cite{Gharib2021COPri} significantly extends the initial ontology with COPri v.2, expanding the concept set from 38 to 63 refined concepts. This version incorporates expert feedback and improves the model’s expressiveness. It introduces key notions such as consent, delegation, trust, and monitoring, which are crucial for modeling complex privacy scenarios in socio-technical systems. In the same it demonstrates its practical utility in detecting nuanced privacy risks.
 
Together, Gharib's works provide a theoretically grounded and practically applicable framework for early-stage privacy requirements engineering, clearly delineating privacy from general security and enabling more precise alignment with PbD principles.

%Conclusion
\paragraph{\textbf{Conclusion for the Data Security and Privacy domain.}} In this domain, ontologies are consistently adopted, with some efforts explicitly publishing them in online repositories, which is less common in other domains. KGs are absent, and semantic logs are not used, showing that the integration of real-time or graph-based data models is not yet developed. A notable exception is one study that integrates NLP/LLM methods with ontological modeling, highlighting an early step toward hybrid symbolic–statistical approaches.

In the subdomain of anonymity, pseudonymity, unlinkability, undetectability, or unobservability, most works rely on static ontologies with OWL DL, but one contribution advances toward a dynamic model with VS+S, integrating NLP for greater automation. Digital rights management shows a stronger focus on dynamic ontologies with applied VS+S systems, including works that provide publicly available ontologies, which enhances reuse and community adoption.

In privacy requirements and data management systems, contributions remain more conservative, relying on static ontologies with OWL DL and focusing mainly on vocabulary semantics (VS) rather than system-level solutions. No advanced reasoning, temporal logic (TL), or fuzzy frameworks (FF) are reported.

%Summary
\definecolor{lightorange}{RGB}{255,245,230}
\begin{tcolorbox}[
  colback=lightorange,   % background color
  colframe=lightorange,  % border color
  boxrule=0pt,           % no visible border
  arc=0mm,               % sharp corners
  left=2mm, right=2mm,   % padding
  top=1mm, bottom=1mm
]
\textbf{Summary:} Table~\ref{tab:domian_digital_rights} shows that the Data Security and Privacy domain maintains a steady reliance on ontologies, with some progress toward dynamic semantics, public availability, and initial steps in hybridization with NLP/LLM. Nevertheless, the domain still lacks broader adoption of knowledge graphs, semantic logs, and formal reasoning frameworks, which limits its ability to address evolving privacy threats and real-time data management challenges.
\end{tcolorbox}

% 4.3 Education and Training
\subsection{Education and Training}
This domain focuses on acquiring the knowledge, skills, and competencies required to protect network and information systems, their users, and impacted individuals from cyber threats. It is composed of 6 subdomains \cite{JRC2021CybersecurityTaxonomy, Nai2019}. We present some ontological works for 2 subdomains according to the inclusion criteria from Section \ref{sec:2.2}:
\begin{inparaenum}[(\bgroup\bfseries 1\egroup)]
\item \emph{Cyber ranges, capture the flag, exercises, simulation platforms, educational/training tools, cybersecurity awareness} that provide hands-on environments and activities designed to develop, assess, and raise skills and awareness in real-world cybersecurity scenarios; and 
\item \emph{Education methodology} emphasizes active, learner-centered approaches that integrate problem-solving, scenario-based learning, and iterative practice. By aligning instructional strategies with real-world threat landscapes, it cultivates both technical proficiency and strategic thinking in cybersecurity education.
\end{inparaenum}

To support a comprehensive analysis of ontology-driven approaches in the Education and Training domain, Table~\ref{tab:domian_education_and_training} presents a comparative overview of relevant research efforts, emphasizing the presence and characteristics of key components such as ontologies, KGs, semantic logic, and formal methods.

%Table
\newcolumntype{C}[1]{>{\centering\arraybackslash}p{#1}}

\begin{table}[h]
\centering
\rowcolors{2}{gray!10}{white}
\begin{tabular}{p{3.8cm} C{1cm} C{0.8cm} C{1.cm} C{1cm} C{1.4cm} C{1.4cm} C{1.5cm} C{1.4cm}}
\toprule
\textbf{Domain, Subdomain and Work} & \textbf{Ontology} & \textbf{KG} & \textbf{Semantic Log} & \textbf{LLM} & \textbf{OWL Type} & \textbf{Ontology Modeling} & \textbf{Semantic Solution} & \textbf{Formal Approach} \\
\midrule

\rowcolor{gray!30} \multicolumn{9}{l}{\textbf{Education and Training}} \\
\midrule

% Cyber ranges, capture the flag exercises, simulation platforms, educational/training tools, cybersecurity awareness
\rowcolor{lightgray}\multicolumn{9}{l}{ \textbf{Cyber ranges, capture the flag exercises, simulation platforms, educational/training tools, cybersecurity awareness}} \\
\midrule
Wen et al.~\cite{Wen2021} & \cellcolor{green!25}\ding{51} & \cellcolor{green!25}\ding{51} & \cellcolor{red!25}\ding{53} & \cellcolor{red!25}\ding{53} & DL & D & VS+S & -- \\

Babayeva et al.~\cite{Babayeva2022} & \cellcolor{green!25}\ding{51} & \cellcolor{red!25}\ding{53}  & \cellcolor{red!25}\ding{53} & \cellcolor{red!25}\ding{53} & -- & S & VS & -- \\

Zacharis et al.~\cite{Zacharis2023} & \cellcolor{green!25}\ding{51} & \cellcolor{green!25}\ding{51}  & \cellcolor{green!25}\ding{51} & \cellcolor{red!25}\ding{53} & -- & D & VS+S & -- \\

% Education methodology
\midrule
\rowcolor{lightgray} \multicolumn{9}{l}{\textbf{Education methodology}} \\
\midrule
Modiba et al. \cite{Modiba2019} & \cellcolor{green!25}\ding{51} & \cellcolor{red!25}\ding{53} & \cellcolor{red!25}\ding{53} & \cellcolor{red!25}\ding{53} & -- & D & VS & -- \\

Agrawal et al. \cite{Agrawal2023} & \cellcolor{green!25}\ding{51} & \cellcolor{green!25}\ding{51} & \cellcolor{green!25}\ding{51} &  \cellcolor{blue!25}\ding{51} NLP & -- & D & VS+S & -- \\

Yamin et al. \cite{Yamin2024} & \cellcolor{green!25}\ding{51} & \cellcolor{green!25}\ding{51} & \cellcolor{green!25}\ding{51} & \cellcolor{green!25}\ding{51} & -- & D & VS+S & -- \\

\midrule
\end{tabular}

\caption{Overview of research efforts in the \textbf{Education and Training} domain.
A checkmark (\ding{51}) on a green background indicates the presence of a component (e.g., \textbf{Ontology}, \textbf{KG}, \textbf{Semantic Log}, or \textbf{LLM}). A cross (\ding{53}) on a red background denotes its absence, while a cross on a purple background indicates that, although no LLM is used, other AI approach are employed (e.g. NLP for Natural Language Processing). The \textbf{OWL Type} column specifies the formal ontology language adopted (e.g., OWL Lite, OWL DL, or OWL Full).
The \textbf{Ontology Modeling} indicates whether the ontology models aspects of the domain: (S) static or (D) dynamic.
The \textbf{Semantic Solution} column distinguishes between ontologies that provide only vocabulary semantics (VS) and those that also define an operational system or application (VS+S). The \textbf{Formal Approach} field identifies whether a reasoning framework is employed, such as logical inference (LI), temporal logic (TL), a fuzzing framework (FF), or enhanced reasoning mechanisms based on temporal logic.}
\label{tab:domian_education_and_training}
\end{table}

%2021 Ontology-Based Scenario Modeling for Cyber Security Exercise
The learning process in cybersecurity involves acquiring the knowledge, skills, and competencies necessary to protect network and information systems, their users, and affected individuals from cyber threats. A prominent example is Locked Shields\footnote{\url{https://ccdcoe.org/locked-shields}}, a large-scale, real-time cyber defence exercise organized by the NATO Cooperative Cyber Defence Centre of Excellence (CCDCOE)\footnote{\url{https://ccdcoe.org/}} in Tallinn. Participants use digital tools to detect, mitigate, and respond to complex cyber incidents under pressure. In contrast, non-technical exercises such as the annual Cyber 9/12 Challenge\footnote{\url{https://www.atlanticcouncil.org/programs/cyber-statecraft-initiative/cyber-912/}} provide a strategic outlook. This desk-based competition invites students worldwide to develop policy responses to fictional cyber crises, emphasizing decision-making over technical execution.

To address the growing need for cybersecurity professionals with practical expertise, ontology-based models have been introduced to systematically structure cybersecurity exercises and cyber range scenarios globally.

% 4.3.1 Cyber ranges, capture the flag exercises, simulation platforms, educational/training tools, cybersecurity awareness
\subsubsection{\textbf{Cyber ranges, capture the flag, exercises, simulation platforms, educational/training tools, cybersecurity awareness}}

% 2021 - Ontology-based scenario modeling for cyber security exercise
An ontology-based model for cyber ranges is introduced by Wen et al.~\cite{Wen2021} to enhance scenario representation and knowledge sharing. The framework is structured into three sub-models: Scenario Information, Scenario Operation, and Security Knowledge, which together capture scenario attributes, event execution, and threat taxonomies. Demonstrated use cases show its effectiveness in modeling context, injects, and attack patterns, while also enabling scenario reuse and supporting the development of a structured scenario repository.

%2022 - Building an ontology for cyber defence exercises
Building on a similar foundation, Babayeva et al.~\cite{Babayeva2022} propose an ontology-based knowledge management system tailored specifically for Cyber Defence Exercises (CDX). While Wen et al. focus on general cyber range scenarios and threat modeling, Babayeva et al. address the unique requirements of CDXs, such as standardizing evaluation data, capturing organizational structures, and supporting training assessments. 

% AI
%2023 AiCEF: an AI-assisted cyber exercise content generation framework using named entity recognition
The AiCEF framework introduces AI-driven automation for generating Cybersecurity Exercise (CSE) content, reducing the manual effort of scenario creation \cite{Zacharis2023}. It leverages the Cyber Exercise Scenario Ontology (CESO) to structure information extracted from cybersecurity articles via Named Entity Recognition (NER). Through a six-step process combining clustering, graph comparison, and synthetic text generation, AiCEF produces technically rich and pedagogically valuable scenarios. Its novelty lies in integrating ML and NLP techniques with ontology-based structuring to improve scenario quality, instructional depth, and development efficiency.

% 4.3.2 Education methodology
\subsubsection{\textbf{Education methodology}}

% **New** - 2019 - An Ontology Based Model for Cyber Security Awareness Education
A formal ontology for structuring the Cybersecurity Awareness (CSA) Body of Knowledge is developed by Mobida et al.~\cite{Modiba2019}. Building on the CURONTO curriculum ontology~\cite{al2013curonto}, it extends the model with classes such as Course, Faculty, Syllabus, Knowledge Areas (KAs), Knowledge Units (KUs), Topics, and Assessment Criteria. The resulting ontology provides a reusable, shareable, and interoperable semantic model for CSA education, facilitating knowledge alignment, curriculum design, and conceptual clarity across institutions.

%2023 - AISecKG: Knowledge Graph Dataset for Cybersecurity Education
AISecKG, introduced by Agrawal et al. \cite{Agrawal2023}, is a novel ontology and annotated dataset for cybersecurity education. It addresses the scarcity of structured, domain-specific resources for novice learners and supports visual, graph-based learning to simplify complex cybersecurity concepts.

% AI - LLMs
% 2024 - Applications of LLMs for Generating Cyber Security Exercise Scenarios
A methodology combining LLMs and ontologies for generating complex cybersecurity exercise scenarios is introduced in \cite{Yamin2024}.
LLMs simulate both known and novel threats, while ontologies ensure semantic consistency, with hallucinations repurposed to create adaptive high-challenge scenarios. Built on the eLuna game framework \cite{eluna}, the approach structures six modules—story, scripts, events, injects, conditions, and infrastructure—formalized through ontologies and aligned with cyber domain knowledge. Two LLMs, supported by RAG and configurable parameters, enable control, diversity, and adaptability, shifting training from static tools to dynamic, evolving threat environments.

%Conclusion
\paragraph{\textbf{Conclusion for the Education and Training domain}} In the Education and Training domain, ontologies are consistently used across all works, but their integration with KGs and semantic logs varies by subdomain. In the area of cyber ranges, capture-the-flag exercises, simulation platforms, training tools, and awareness, studies rely mainly on static or dynamic ontologies with OWL DL where specified, and most deliver VS+S solutions that directly support educational applications. However, no advanced reasoning frameworks are employed. In the education methodology subdomain, contributions demonstrate a stronger shift toward dynamic ontologies and applied VS+S solutions, with one study integrating NLP/LLM methods alongside ontologies, an early but important step toward hybrid symbolic–statistical approaches. Ontology reuse also appears in this subdomain, though formal reasoning remains absent.

\definecolor{lightorange}{RGB}{255,245,230}
\begin{tcolorbox}[
  colback=lightorange,   % background color
  colframe=lightorange,  % border color
  boxrule=0pt,           % no visible border
  arc=0mm,               % sharp corners
  left=2mm, right=2mm,   % padding
  top=1mm, bottom=1mm
]
\textbf{Summary:} Table~\ref{tab:domian_education_and_training} shows that the Education and Training domain places clear emphasis on practical system support (VS+S) for training and methodology, alongside emerging progress in dynamic semantics, ontology reuse, and LLM integration. However, the limited adoption of KGs, semantic logging, and formal reasoning frameworks continues to constrain its ability to capture evolving learning contexts and to support fully adaptive, intelligent training environments.
\end{tcolorbox}

% 4 - Incident Handling and Digital Forensics
\subsection{Incident Handling and Digital Forensics}

This domain refers to theories, techniques, tools and processes for the identification, collection, acquisition, and preservation of digital evidence \cite{JRC2021CybersecurityTaxonomy, Nai2019}. It includes incident response activities such as analysis, documentation, attribution, forecasting, and coordinated reporting. The field also addresses vulnerability analysis, forensic workflows, case studies, policy issues, and resilience against anti-forensics and malware tactics. Out of the 10 domains, we collected relevant works for 5 subdomains selected according to the criteria outlined in Section~\ref{sec:2.2} as described after: 
\begin{inparaenum}[(\bgroup\bfseries 1\egroup)]
\item \emph{Anti-forensics and malware analytics} focuses on techniques used to detect, analyze, and counteract efforts to conceal, obfuscate, or manipulate digital evidence, as well as the behavioral analysis of malicious software;
\item \emph{Coordination and information sharing in the context of cross-border/organizational incidents} covers the protocols and frameworks for effective collaboration and real-time information exchange across jurisdictions and organizational boundaries during cybersecurity incidents;
\item \emph{Digital forensics processes workflow models} defines structured methodologies and lifecycle models for conducting digital forensic investigations, from evidence acquisition to reporting and archiving;
\item \emph{Incident analysis, communication, documentation, forecasting, response, and reporting}
encompasses the systematic evaluation of security events and the execution of response strategies, along with transparent documentation, stakeholder communication, and predictive analysis; and 
\item \emph{Vulnerability analysis and response} involves identifying, assessing, prioritizing, and mitigating security vulnerabilities to reduce risk and strengthen system resilience against exploitation.
\end{inparaenum}

To facilitate a thorough examination of ontology-based approaches within the Incident Handling and Digital Forensics domain, Table~\ref{tab:domian_incident_handling_and_digital_forensics} provides a comparative summary of pertinent research works, highlighting the inclusion and features of core elements such as ontologies, KGs, semantic reasoning, and formal techniques.

% Table
\newcolumntype{C}[1]{>{\centering\arraybackslash}p{#1}}

\begin{table}[h]
\centering
\rowcolors{2}{gray!10}{white}
\begin{tabular}{p{3.8cm} C{1cm} C{0.8cm} C{1.cm} C{1cm} C{1.4cm} C{1.4cm} C{1.5cm} C{1.4cm}}
\toprule
\textbf{Domain, Subdomain and Work} & \textbf{Ontology} & \textbf{KG} & \textbf{Semantic Log} & \textbf{LLM} & \textbf{OWL Type} & \textbf{Ontology Modeling} & \textbf{Semantic Solution} & \textbf{Formal Approach} \\
\midrule

\rowcolor{gray!30}\multicolumn{9}{l}{ \textbf{Incident Handling and Digital Forensics}} \\
\midrule

% Cyber ranges, capture the flag exercises, simulation platforms, educational/training tools, cybersecurity awareness
\rowcolor{lightgray}\multicolumn{9}{l}{ \textbf{Anti-forensics and malware analytics}} \\
\midrule
Tafazzoli et al. \cite{Tafazzoli2008} & \cellcolor{green!25}\ding{51} & \cellcolor{red!25}\ding{53}  & \cellcolor{red!25}\ding{53} & \cellcolor{red!25}\ding{53} & DL & S & VS & FF \\

Obrst et al. \cite{Obrst2012} & \cellcolor{green!25}\ding{51} & \cellcolor{red!25}\ding{53}  & \cellcolor{red!25}\ding{53} & \cellcolor{red!25}\ding{53} & DL & S & VS+S & TL \\

Dimitriadis et al. \cite{Dimitriadis2022} & \cellcolor{green!25}\ding{51} & \cellcolor{green!25}\ding{51} & \cellcolor{green!25}\ding{51} & \cellcolor{blue!25}\ding{51} ML & -- & D & VS+S & TL \\

Nazoksara et al. \cite{Nazoksara2025}  & \cellcolor{green!25}\ding{51} &  \cellcolor{red!25}\ding{53} &  \cellcolor{red!25}\ding{53} & \cellcolor{blue!25}\ding{51} DL & -- & D & VS+S & -- \\
\midrule

%Coordination and information sharing in the context of cross-border/organizational incidents
\rowcolor{lightgray}\multicolumn{9}{l}{ \textbf{ Coordination and information sharing in the context of cross-border/organizational incidents}} \\
\midrule

Takahashi et al. \cite{Takahashi2015} & \cellcolor{green!25}\ding{51} &  \cellcolor{red!25}\ding{53} &  \cellcolor{red!25}\ding{53} &  \cellcolor{red!25}\ding{53} & -- & S & VS & -- \\

Oltramari et al. \cite{Oltramari2015} & \cellcolor{green!25}\ding{51} & \cellcolor{green!25}\ding{51} & \cellcolor{green!25}\ding{51} & \cellcolor{red!25}\ding{53} & DL & D & VS+S & -- \\

Onwubiko et al. \cite{Onwubiko2018} & \cellcolor{green!60}\ding{79} \cite{conwubiko_cocoa_2025} & \cellcolor{green!25}\ding{51} & \cellcolor{green!25}\ding{51} & \cellcolor{red!25}\ding{53} & -- & D & VS+S & -- \\
\midrule

%Incident analysis, communication, documentation, forecasting, response, and reporting
\rowcolor{lightgray} \multicolumn{9}{l}{ \textbf{Incident analysis, communication, documentation, forecasting, response, and reporting}} \\
\midrule

Chockalingam et al. \cite{Chockalingam2022} & \cellcolor{green!25}\ding{51} & \cellcolor{red!25}\ding{53} & \cellcolor{red!25}\ding{53} & \cellcolor{red!25}\ding{53} & DL & S & VS+S & -- \\

Posea et al. \cite{Posea2022} & \cellcolor{green!25}\ding{51} & \cellcolor{red!25}\ding{53} & \cellcolor{red!25}\ding{53} & \cellcolor{red!25}\ding{53} & DL & S & VS & -- \\

Ben-Shimol et al. \cite{BenShimol2024} & \cellcolor{green!25}\ding{51} & \cellcolor{red!25}\ding{53} & \cellcolor{red!25}\ding{53} & \cellcolor{red!25}\ding{53} & DL & S & VS & -- \\

%Vulnerability analysis and response
\midrule
\rowcolor{lightgray}\multicolumn{9}{l}{ \textbf{Vulnerability analysis and response}} \\
\midrule

Wang et al. \cite{Wang2009} & \cellcolor{green!25}\ding{51} & \cellcolor{red!25}\ding{53} & \cellcolor{red!25}\ding{53} & \cellcolor{red!25}\ding{53} & -- & S & VS+S & -- \\

Romilla Syed \cite{Syed2020} & \cellcolor{green!25}\ding{51} & \cellcolor{red!25}\ding{53} & \cellcolor{red!25}\ding{53} & \cellcolor{red!25}\ding{53} & -- & S & VS+S & -- \\

Lindroth \cite{Lindroth2022} & \cellcolor{green!25}\ding{51} & \cellcolor{red!25}\ding{53} & \cellcolor{red!25}\ding{53} & \cellcolor{red!25}\ding{53} & -- & S & VS+S & -- \\

Hu et al. \cite{Hu2022} & \cellcolor{green!25}\ding{51} & \cellcolor{red!25}\ding{53} & \cellcolor{red!25}\ding{53} & \cellcolor{red!25}\ding{53} & DL & S & VS+S & -- \\

\midrule
\end{tabular}

\caption{Overview of research efforts in the \textbf{Incident Handling and Digital Forensics} domain.
A checkmark (\ding{51}) on a green background indicates the presence of a component (e.g., \textbf{Ontology}, \textbf{Knowledge Graph}, \textbf{Semantic Log}, or \textbf{LLM}). A cross (\ding{53}) on a red background denotes its absence, while a cross on a purple background indicates that, although no LLM is used, other AI approach are employed (e.g. ML for Machine Learning and DL for Deep Learning). A star (\ding{79}) on green background marks ontologies that are publicly available in online repositories. The \textbf{OWL Type} column specifies the formal ontology language adopted (e.g., OWL Lite, OWL DL, or OWL Full).
The \textbf{Ontology Modeling} indicates whether the ontology models aspects of the domain: (S) static or (D) dynamic.
The \textbf{Semantic Solution} column distinguishes between ontologies that provide only vocabulary semantics (VS) and those that also define an operational system or application (VS+S). The \textbf{Formal Approach} field identifies whether a reasoning framework is employed, such as logical inference (LI), temporal logic (TL), a fuzzing framework (FF), or enhanced reasoning mechanisms based on temporal logic.}
\label{tab:domain_incident_handling_and_digital_forensics}
\end{table}

% 4.4.1 Anti-forensics and malware analytics.
\subsubsection{\textbf{Anti-forensics and malware analytics}} 
%2008 - Malware fuzzy ontology for semantic web
A malware ontology for mapping semantic relationships between malware concepts is proposed by Tafazzoli et al.~\cite{Tafazzoli2008}. Designed to support semantic-based search engines for incident management and Computer Emergency Response Team (CERT) portals, it addresses overlapping malware characteristics by applying fuzzy logic to model unclear concept boundaries. Relationships are classified into five strength levels—from very weak to very good—each assigned a weight. Queries on specific concepts are evaluated against these weighted relationships, with searches dynamically expanding to related concepts based on query context and connection strength. The work also envisions extending the ontology into a broader Cyber ontology, incorporating relevant standards and utility domains.

%2012 Developing an Ontology of the Cyber Security Domain
A trade study for evolving a cyber ontology from a malware-focused model is presented, emphasizing reuse and integration of standards such as MAEC\footnote{http://maec.mitre.org} , OpenIOC\footnote{https://cloud.google.com/blog/topics/threat-intelligence/openioc-basics/}, CybOX\footnote{https://cybox.mitre.org/about/}, STIX\footnote{http://measurablesecurity.mitre.org/docs/STIX-Whitepaper.pdf}, and CAPEC\footnote{http://capec.mitre.org/}.
The architecture is modular, combining upper, mid-level, domain-specific, and utility ontologies to improve scalability and interoperability.
Foundational ontologies including DOLCE \cite{dolceLOA2025}, BFO \cite{obrst2012developing}, and SUMO \cite{niles2001standard} ensure semantic consistency and cross-domain alignment. Utility ontologies cover cross-cutting concerns like time, space, agents, and events, extending beyond Tafazzoli’s narrower, search-focused framework.

%2022 Fronesis: Digital Forensics-Based Early Detection of Ongoing Cyber-Attacks
Fronesis, introduced in~\cite{Dimitriadis2022}, is a system for early cyber-attack detection through ontological reasoning. It integrates the MITRE ATT\&CK\footnote{https://attack.mitre.org/}
 framework and the Cyber Kill Chain\footnote{https://www.lockheedmartin.com/en-us/capabilities/cyber/cyber-kill-chain.html}
 model, adopting a practical, operational approach to cyber defense. By mapping digital artifacts to ATT\&CK tactics and CKC phases, Fronesis enables evidence-based, real-time reconstruction of attack paths. Its integration of semantic reasoning with forensic data supports dynamic situational awareness and live threat detection.

% ML
%2024 - SAutoIDS: A Semantic Autonomous Intrusion Detection System Based on Cellular Deep Learning and Ontology for Malware Detection in cloud computing
SAutoIDS, presented by Nazoksara et al.~\cite{Nazoksara2025}, is a cloud-based intrusion detection system that combines ontological feature selection with deep learning. Unlike Tafazzoli et al., who apply fuzzy ontologies for semantic search, SAutoIDS leverages ontologies to abstract feature spaces for malware detection, prioritizing classification accuracy in cloud environments rather than real-time correlation. Compared to Fronesis, it operates at the application layer. Its novelty lies in integrating semantic abstraction with the Group Method of Data Handling Deep Neural Network (GMDH-DNN) and cellular automata, offering a hybrid approach to intrusion detection.

% 4.4.2 Coordination and information sharing in the context of cross-border/organizational incidents
\subsubsection{\textbf{Coordination and information sharing in the context of cross-border/organizational incidents}}

%2014 Reference Ontology for Cybersecurity Operational Information
An ontology for cybersecurity operational information is proposed in~\cite{Takahashi2015} to address the increasing need for a unified framework that organizes and orchestrates diverse industry specifications for information exchange. Developed collaboratively with cybersecurity organizations, the ontology structures cybersecurity data, integrates existing specifications, and improves interoperability in operational contexts. It defines three key operation domains—IT Asset Management, Incident Handling, and Knowledge Accumulation—along with essential roles, databases, and knowledge bases that store and manage critical information. 

%2015 Computational ontology of network operations
CRATELO, introduced by  Oltramar et al. ~\cite{Oltramari2015}, is a computational framework designed to enhance cyber defense through semantic integration in network operations. Emphasizing situational awareness and dynamic reasoning in secure environments, it contrasts with Takahashi et al.’s focus on structuring cybersecurity information around roles and standards. CRATELO adopts a modular, three-tiered architecture grounded in formal knowledge representation, integrating OSCO, SECCO~\cite{oltramari2014building}, and foundational alignment with DOLCE to ensure interoperability between operational data and conceptual reasoning. The framework also emphasizes intrusion detection and the application of cyber semantics for proactive, intelligence-driven defense.

%2018 CoCoa: An Ontology for Cybersecurity Operations Centre Analysis Process
Onwubiko et al~\cite{Onwubiko2018}, introduce CoCoa as an ontology tailored to model cyber analysis processes in Cybersecurity Operations Centers (CSOCs). Aligned with the NIST Cybersecurity Framework lifecycle—Identify, Protect, Detect, Respond, and Recover—it emphasizes real-time workflows and analytical tasks within operational environments. Unlike Takahashi et al., who focus on standardizing cybersecurity information exchange, CoCoa models process-oriented activities; and in contrast to CRATELO, which prioritizes semantic reasoning, it provides a lightweight ontology grounded in CSOC practices. By constructing a knowledge graph-based model, CoCoa supports situational monitoring and traceability, bridging abstract ontology design with actionable cyber threat response.

% 4.4.3 Incident analysis, communication, documentation, forecasting, response, and reporting
\subsubsection{\textbf{Incident analysis, communication, documentation, forecasting, response, and reporting}}

%2020 An ontology for effective security incident management
An ontology for structured security incident management is proposed to address gaps in the mid- and post-incident phases \cite{Chockalingam2022}.
It formalizes concepts such as actors, assets, attack vectors, detection, forensics, and response, with semantic relations enabling reasoning and automation. Evaluated for logical consistency and applied to the Colonial Pipeline ransomware case\footnote{https://www.cisa.gov/news-events/news/attack-colonial-pipeline-what-weve-learned-what-weve-done-over-past-two-years}, it improves incident understanding and response coordination. The ontology fosters knowledge reuse and provides a foundation for future AI/ML-driven security applications.

%2022 Towards unified european cyber incident and crisis management ontology
A Unified European Cyber Incident and Crisis Management Ontology is proposed to improve interoperability among analysts and institutions across borders \cite{Posea2022}. Unlike existing taxonomies that emphasize only technical aspects, it introduces the concept of “Observation” to structure evidence and metadata for consistent reporting. The ontology is implemented in OWL and reuses resources such as the JRC Taxonomy \cite{Fovino2019}, UCO \cite{Syed2016}, and ENISA Reference Incident Classification Taxonomy \cite{ENISA2018}. It extends the scope beyond organizational-level ontologies, supporting cross-sector and cross-border coordination.
This enables real-time situational awareness and unified crisis communication at the European level.

%2024 Observability and Incident Response in Managed Serverless Environments Using Ontology-Based Log Monitoring
An ontology-based approach is proposed by Ben-Shimol et al. \cite{BenShimol2024} to enhance observability and incident response in managed serverless environments. The architecture introduces a three-layer model that transforms heterogeneous cloud-native logs into a unified activity knowledge graph. This graph supports an incident response dashboard for contextual alert analysis and a criticality of asset framework for risk-based prioritization. Unlike works focusing on organizational phases or cross-border harmonization, the approach targets cloud-native, real-time environments requiring system-level adaptability.

% 4.4.4 Vulnerability analysis and response
\subsubsection{\textbf{Vulnerability analysis and response}}

%2009 Ontology-based security assessment for software products
Wang et al. \cite{Wang2009} propose the Ontology for Vulnerability Management (OVM), a structured and semantically rich framework that unifies multiple standards as CVSS, CVE, CWE, CPE, and CAPEC to holistically assess software security. Their method moves beyond evaluating isolated vulnerabilities by grouping them into environmental cases based on contextual attributes such as access vector and authentication level, thereby computing an aggregated trust metric that reflects the software’s security posture in its operational environment. 

%2020 Cybersecurity vulnerability management: A conceptual ontology and cyber intelligence alert system
This system-level, trust-oriented approach contrasts with Romilla Syed \cite{Syed2020} Cybersecurity Vulnerability Ontology (CVO), which focuses on formal knowledge representation and automated alert generation via the Cyber Intelligence Alert (CIA) system. While both integrate standard vulnerability taxonomies, Romilla Syed extend applicability by incorporating real-time threat intelligence from social media, introducing a dynamic, data-driven component absent in Wang et al. model.

%2022 Cybersecurity Ontology-The relationship between vulnerabilities, standards, legal and regulatory requirements
Lindroth \cite{Lindroth2022}, in turn, contributes a security ontology centered on regulatory alignment, mapping relationships between vulnerabilities, controls, and compliance requirements. Unlike Wang and Syed, whose works focus on operational management and automation, Lindroth emphasizes governance and legal interpretation, offering a tool for practitioners and stakeholders navigating complex compliance landscapes.

%2022 Software security vulnerability patterns based on ontology
The conceptual and lifecycle aspects of vulnerabilities within DevSecOps are addressed through a multi-layered ontology proposed by Hu et al. \cite{Hu2022}, which clarifies distinctions between coding errors and security flaws. This model enhances analysis across the development pipeline—from risk modeling to penetration testing—positioning itself closer to Wang et al.’s goal of systemic security evaluation, but with a stronger emphasis on development practices and vulnerability characterization.

%Conclusion
\paragraph{\textbf{Conclusion for the Incident Handling and Digital Forensics domain}} In the Incident Handling and Digital Forensics domain, ontologies are a common foundation across all subdomains, though their integration with KGs, semantic logs, and ML/LLM is limited and fragmented. In anti-forensics and malware analytics, most works employ static DL ontologies, with one contribution extending toward dynamic semantics, ML integration, and temporal logic reasoning, marking an early hybrid approach. Coordination and information sharing shows stronger maturity, with dynamic ontology models, some knowledge graph use, and the only clear example of ontology reuse, although formal reasoning is still absent. 
The subdomain of incident analysis, communication, documentation, forecasting, response, and reporting remains conservative, relying on static DL ontologies and delivering either basic vocabularies (VS) or system-supported semantics (VS+S), but without KG integration or advanced reasoning. In contrast, vulnerability analysis and response emphasizes applied VS+S solutions, offering practical support but relying on static models and avoiding reuse or formal frameworks.

\definecolor{lightorange}{RGB}{255,245,230}
\begin{tcolorbox}[
  colback=lightorange,   % background color
  colframe=lightorange,  % border color
  boxrule=0pt,           % no visible border
  arc=0mm,               % sharp corners
  left=2mm, right=2mm,   % padding
  top=1mm, bottom=1mm
]
\textbf{Summary:} Table \ref{tab:domain_incident_handling_and_digital_forensics} highlights this domain’s strong reliance on ontologies, with emerging advances in dynamic semantics, ML integration, and reuse, particularly in areas such as anti-forensics and coordination. However, the limited adoption of knowledge graphs, semantic logging, and advanced reasoning frameworks constrains its capacity to fully enable dynamic, real-time incident handling and forensic analysis.
\end{tcolorbox}

% 5 Network and Distributed Systems
\subsection{Network and Distributed Systems}
This domain focuses on safeguarding data and communication across networked and distributed systems. It addresses the integrity, confidentiality, availability, and authenticity of messages, as well as secure coordination and computation among distributed components. It includes protection of hardware, software, communication protocols, and message authentication mechanisms \cite{JRC2021CybersecurityTaxonomy, Nai2019}. The domain comprises 15 subdomains, for which our work gathers existing studies and ontologies for 3 selected according to the criteria outlined in Section~\ref{sec:2.2} and detailed right away:
\begin{inparaenum}[(\bgroup\bfseries 1\egroup)]
\item \emph{Distributed consensus techniques} are protocols that ensure agreement among distributed nodes, even in the presence of faults;
\item \emph{Fault-tolerant models} are system designs that ensure continued operation despite component failures; and
\item \emph{Requirements for network security} are core safeguards—such as authentication, integrity, and availability—that protect networked systems.
\end{inparaenum}
Table~\ref{tab:domian_incident_handling_and_digital_forensics} presents a comparative overview of research efforts within the Network and Distributed Systems domain, emphasizing the presence and features of key elements such as ontologies, KGs, semantic reasoning, and formal methods.

% Table
\newcolumntype{C}[1]{>{\centering\arraybackslash}p{#1}}

\begin{table}[h]
\centering
\rowcolors{2}{gray!10}{white}
\begin{tabular}{p{3.8cm} C{1cm} C{0.8cm} C{1.cm} C{1cm} C{1.4cm} C{1.4cm} C{1.5cm} C{1.4cm}}
\toprule
\textbf{Domain, Subdomain and Work} & \textbf{Ontology} & \textbf{KG} & \textbf{Semantic Log} & \textbf{LLM} & \textbf{OWL Type} & \textbf{Ontology Modeling} & \textbf{Semantic Solution} & \textbf{Formal Approach} \\
\midrule
\rowcolor{gray!30} \multicolumn{9}{l}{\textbf{Network and Distributed Systems}} \\
\midrule

% Distributed consensus techniques
\rowcolor{lightgray}\multicolumn{9}{l}{ \textbf{Distributed consensus techniques}} \\
\midrule
Dastjerdi et al. \cite{Dastjerdi2015} & \cellcolor{green!25}\ding{51} & \cellcolor{red!25}\ding{53}  & \cellcolor{red!25}\ding{53} & \cellcolor{red!25}\ding{53} & DL & D & VS+S & -- \\

Bonyadi \cite{Bonyadi2024} & \cellcolor{green!25}\ding{51} & \cellcolor{red!25}\ding{53}  & \cellcolor{red!25}\ding{53} & \cellcolor{red!25}\ding{53} & -- & S & VS & -- \\

% Fault-tolerant models
\midrule
\rowcolor{lightgray}\multicolumn{9}{l}{ \textbf{Fault-tolerant models}} \\
\midrule

Jaquete\cite{jaquetemeta} & \cellcolor{green!25}\ding{51} & \cellcolor{red!25}\ding{53}  & \cellcolor{red!25}\ding{53} & \cellcolor{red!25}\ding{53} & -- & D & VS+S & FF \\

 et al. \cite{Diao2022} & \cellcolor{green!25}\ding{51} & \cellcolor{red!25}\ding{53}  & \cellcolor{red!25}\ding{53} & \cellcolor{red!25}\ding{53} & DL & S & VS+S & -- \\

\midrule
\rowcolor{lightgray} \multicolumn{9}{l}{\textbf{Requirements for network security}} \\
\midrule

Bhandari et al. \cite{Bhandari2014} & \cellcolor{green!25}\ding{51} & \cellcolor{red!25}\ding{53}  & \cellcolor{green!25}\ding{51} & \cellcolor{red!25}\ding{53} & DL & D & VS+S & -- \\

Velasco et al. \cite{Velasco2017} & \cellcolor{green!25}\ding{51} & \cellcolor{red!25}\ding{53}  & \cellcolor{red!25}\ding{53} & \cellcolor{red!25}\ding{53} & -- & S & VS & -- \\

\midrule
\end{tabular}
\caption{Overview of research efforts in the \textbf{Network and Distributed Systems} domain.
A checkmark (\ding{51}) on a green background indicates the presence of a component (e.g., \textbf{Ontology}, \textbf{Knowledge Graph}, \textbf{Semantic Log}, or \textbf{LLM}). A cross (\ding{53}) on a red background denotes its absence. The \textbf{OWL Type} column specifies the formal ontology language adopted (e.g., OWL Lite, OWL DL, or OWL Full).
The \textbf{Ontology Modeling} indicates whether the ontology models aspects of the domain: (S) static or (D) dynamic.
The \textbf{Semantic Solution} column distinguishes between ontologies that provide only vocabulary semantics (VS) and those that also define an operational system or application (VS+S). The \textbf{Formal Approach} field identifies whether a reasoning framework is employed, such as logical inference (LI), temporal logic (TL), a fuzzing framework (FF), or enhanced reasoning mechanisms based on temporal logic.}
\label{tab:domian_incident_handling_and_digital_forensics}
\end{table}

% 4.5.1 Distributed consensus techniques
\subsubsection{\textbf{Distributed consensus techniques}}

%2015 - On Application of Ontology and Consensus Theory to Human-Centric IoT: an Emergency Management Case Study
A framework integrating a domain ontology with consensus theory is proposed to enhance emergency response in Human-Centric IoT systems \cite{Dastjerdi2015}. Developed in WSMO/WSML, it reuses the SSN Ontology\footnote{https://www.w3.org/TR/vocab-ssn/} and a disaster management model by Kruchten et al., capturing disaster types, expert skills, resources, and locations. Ontology-based reasoning supports expert discovery and coordination in distress scenarios, demonstrated through a case study. A grounding module transforms semi-structured data into semantic formats, enabling interoperability and effective resource matching.

%2024 - Toward a Unified, Hierarchical Ontology of Consensus Proofs in Autonomous Distributed Computing
HOPSCA is a hierarchical ontology for formalizing and unifying consensus proofs across distributed systems \cite{Bonyadi2024}. It maps core security properties—termination, validity, and agreement—to system components, enabling consistent modeling of consensus algorithms. Unlike application-specific approaches such as Dastjerdi et al.’s focus on IoT emergency response, HOPSCA emphasizes system-agnostic formalization and standardized verification. The framework ensures backward and forward compatibility in proof structures, facilitating comparison across protocols and advancing the theoretical foundations of consensus.

% 4.5.2 Fault-tolerant models
\subsubsection{\textbf{Fault tolerant models}}

%2014 - A Fault Fuzzy-To overcome thislogy for Large Scale Fault-tolerant Wireless Sensor Networks
A framework that integrates a domain ontology with consensus theory is proposed to enhance emergency response in Human-Centric IoT (HC-IoT) systems. Building on the SSN Ontology and Kruchten et al.’s disaster management model, the ontology formalizes core concepts such as disaster types, expert skills, resources, and locations, thereby ensuring semantic interoperability. Its effectiveness is illustrated through a case study where ontology-based reasoning supports expert discovery and coordination in a distress scenario, while a grounding module translates semi-structured data into semantic formats to enable accurate matching \cite{dastjerdi2015application}.

%2022 - An ontology-based fault generation and fault propagation analysis approach for safetycritical computer systems at the design stage
An ontology-based framework is proposed for analyzing fault generation and propagation in safety-critical systems during the design phase \cite{Diao2022}. Unlike Jaquete’s focus on internal ontology fault detection, this approach models external system faults across hardware and software layers. Using OWL and SWRL, it defines fault types, triggers, and propagation mechanisms, enabling automated fault scenario generation and inference-driven impact analysis. The framework evaluates both single and concurrent faults, supporting early-stage risk assessment and design-time safety validation. 

% 4.5.3 Requirements for network security
\subsubsection{\textbf{Requirements for network security}}

%2014 - Ontology based approach for perception of network security state
Bhandari et al. \cite{Bhandari2014} propose an ontology-based framework for addressing evolving security challenges in distributed IoT environments. Ontologies model threats, countermeasures, and core security requirements
like authentication, confidentiality, and access control.
The approach enables policy definition, threat reasoning, and adaptive decisions. Security is reinforced through key management and encryption techniques. Ontologies support real-time monitoring and anomaly detection. The framework offers a scalable, systematic solution for complex network security.

%2017 - Ontologies for Network Security and Future Challenges
A broader critical review of existing network security ontologies emphasizes limitations such as narrow scope and insufficient validation, as discussed by Velasco et al. \cite{Velasco2017}. This perspective contrasts with the more application-driven and structured framework proposed by Bhandari et al., which focuses on real-time decision-making in heterogeneous environments such as IoT. The novelty of Velasco et al.’s work lies in its holistic classification methodology and its call for unified, formally validated ontologies that encompass a wider range of security aspects—including policies and tools—dimensions often underexplored in earlier efforts.

%Conclusion
\paragraph{\textbf{Conclusion for the Network and Distributed Systems domain.}} In this domain, ontologies are employed across all contributions, but the use of KGs, semantic logs, and ML/LLM methods is entirely absent, indicating a strong reliance on symbolic modeling. For distributed consensus techniques, works adopt OWL DL with both static and dynamic ontologies, producing VS+S solutions that support operational use, though without advanced reasoning. In the subdomain of fault-tolerant models, ontologies are also used, with contributions focused on VS+S solutions. One study employs a FF, making it the only formal reasoning effort in this domain, although no reuse or openness of ontologies is reported. For requirements for network security, works remain mostly conservative: ontologies are present, with one study using OWL DL and dynamic semantics for a VS+S solution, while the other is limited to static vocabulary semantics (VS). Again, no integration with KG, semantic logs, or hybrid AI methods is observed.

\definecolor{lightorange}{RGB}{255,245,230}
\begin{tcolorbox}[
  colback=lightorange,   % background color
  colframe=lightorange,  % border color
  boxrule=0pt,           % no visible border
  arc=0mm,               % sharp corners
  left=2mm, right=2mm,   % padding
  top=1mm, bottom=1mm
]
\textbf{Summary:} Table~\ref{tab:domian_incident_handling_and_digital_forensics} indicates that this domain consistently employs ontologies with an applied orientation (VS+S) across several contributions. Nonetheless, advances in dynamic semantics, formal reasoning, and hybrid approaches remain limited and fragmented. The lack of KGs adoption, log integration, and statistical/LLM methods constrains its ability to address evolving distributed environments and to support network resilience in real-world scenarios.
\end{tcolorbox}

% 6 Security Management and Governance
\subsection{Security Management and Governance}
This domain covers governance and management strategies, methodologies, and tools aimed at safeguarding the core security properties of information: confidentiality, integrity and availability, along with authenticity, accountability, and non-repudiation \cite{JRC2021CybersecurityTaxonomy, Nai2019}. It is organized into 15 subdomains. In our work, we focus on three of these, collecting existing studies and ontologies that meet the criteria defined in Section~\ref{sec:2.2}, as detailed below:
\begin{inparaenum}[(\bgroup\bfseries 1\egroup)]
\item \emph{Attack modeling, techniques, and countermeasures} focuses on representing and analyzing how attacks are executed, identifying their underlying techniques, and designing strategies to detect, prevent, or mitigate them;
\item \emph{Risk management, including modeling, assessment, analysis and mitigation} is a structured process for identifying, evaluating, and reducing security risks to ensure the protection of critical information assets; and 
\item \emph{Threats and vulnerability modeling} involves the identification and representation of potential threat actors, attack vectors, and system weaknesses to support proactive security planning and risk mitigation.
\end{inparaenum}
Table~\ref{tab:domian_security_management_and_governance} presents a comparative overview of research efforts within the Security Management and Governance domain, emphasizing the presence and features of key elements such as ontologies, KGs, semantic reasoning, and formal methods.

% Table
\newcolumntype{C}[1]{>{\centering\arraybackslash}p{#1}}

\begin{table}[h]
\centering
\rowcolors{2}{gray!10}{white}
\begin{tabular}{p{3.8cm} C{1cm} C{0.8cm} C{1.cm} C{1cm} C{1.4cm} C{1.4cm} C{1.5cm} C{1.4cm}}
\toprule
\textbf{Domain, Subdomain and Work} & \textbf{Ontology} & \textbf{KG} & \textbf{Semantic Log} & \textbf{LLM} & \textbf{OWL Type} & \textbf{Ontology Modeling} & \textbf{Semantic Solution} & \textbf{Formal Approach} \\
\midrule

\rowcolor{gray!30}\multicolumn{9}{l}{ \textbf{Security Management and Governance}} \\
\midrule

\rowcolor{lightgray} \multicolumn{9}{l}{\textbf{Attack modeling, techniques, and countermeasures}} \\
\midrule
Ansarinia et al. \cite{ansarinia2012ontology} & \cellcolor{green!25}\ding{51} & \cellcolor{green!25}\ding{51} & \cellcolor{green!25}\ding{51} & \cellcolor{red!25}\ding{53} & DL & S & VS+S & -- \\

Luh et al. \cite{luh2016taon} & \cellcolor{green!25}\ding{51} & \cellcolor{red!25}\ding{53}  & \cellcolor{green!25}\ding{51} & \cellcolor{red!25}\ding{53} & DL & D & VS+S & -- \\

Huang et al. \cite{huang2022building} & \cellcolor{green!25}\ding{51} & \cellcolor{red!25}\ding{53}  & \cellcolor{red!25}\ding{53} & \cellcolor{blue!25}\ding{51} NLP & -- & S & VS+S & -- \\

\midrule
\rowcolor{lightgray}\multicolumn{9}{l}{ \textbf{Risk management, including modeling, assessment, analysis and mitigation}} \\
\midrule

Sales et al. \cite{sales2018common} & \cellcolor{green!25}\ding{51} & \cellcolor{red!25}\ding{53}  & \cellcolor{red!25}\ding{53} & \cellcolor{red!25}\ding{53} & -- & D & VS & TL \\

Vega et al. \cite{vega2019ontology} & \cellcolor{green!25}\ding{51} & \cellcolor{red!25}\ding{53}  & \cellcolor{red!25}\ding{53} & \cellcolor{red!25}\ding{53} & DL & D & VS+S & -- \\

Arogundade et al. \cite{arogundade2020ontology} & \cellcolor{green!25}\ding{51} & \cellcolor{red!25}\ding{53}  & \cellcolor{red!25}\ding{53} & \cellcolor{red!25}\ding{53} & DL & S & VS+S & TL \\

Oliveira et al. \cite{oliveira2022ontology} & \cellcolor{green!60}\ding{79} \cite{unibz_security_ontology_2025} & \cellcolor{red!25}\ding{53}  & \cellcolor{red!25}\ding{53} & \cellcolor{red!25}\ding{53} & DL & S & VS+S & TL \\

Sánchez-Zas et al. \cite{sanchez2023ontology} & \cellcolor{green!25}\ding{51} & \cellcolor{red!25}\ding{53}  & \cellcolor{green!25}\ding{51} & \cellcolor{red!25}\ding{53} & DL & D & VS+S & TL \\

Oliveira et al. \cite{oliveira2025toward} & \cellcolor{green!25}\ding{51} & \cellcolor{red!25}\ding{53}  & \cellcolor{red!25}\ding{53} & \cellcolor{red!25}\ding{53} & DL & D & VS+S & -- \\

\midrule
\rowcolor{lightgray} \multicolumn{9}{l}{\textbf{Threats and vulnerability modeling}} \\
\midrule

Manzoor et al. \cite{manzoor2019threat} & \cellcolor{green!25}\ding{51} & \cellcolor{red!25}\ding{53}  & \cellcolor{red!25}\ding{53} & \cellcolor{red!25}\ding{53} & -- & S & VS+S & -- \\

Valja et al. \cite{valja2020automating} & \cellcolor{green!25}\ding{51} & \cellcolor{red!25}\ding{53}  & \cellcolor{red!25}\ding{53} & \cellcolor{red!25}\ding{53} & -- & S & VS+S & -- \\

Rosa et al. \cite{de2022threma} & \cellcolor{green!25}\ding{51} & \cellcolor{red!25}\ding{53}  & \cellcolor{red!25}\ding{53} & \cellcolor{red!25}\ding{53} & DL & S & VS+S & -- \\

Compierchio \cite{compierchio2024ontology} & \cellcolor{green!25}\ding{51} & \cellcolor{red!25}\ding{53}  & \cellcolor{red!25}\ding{53} & \cellcolor{red!25}\ding{53} & -- & S & VS+S & -- \\
\midrule
\end{tabular}

\caption{Overview of research efforts in the \textbf{Security Management and Governance} domain. A checkmark (\ding{51}) on a green background indicates the presence of a component (e.g., \textbf{Ontology}, \textbf{Knowledge Graph}, \textbf{Semantic Log}, or \textbf{LLM}). A cross (\ding{53}) on a red background denotes its absence, while a cross on a purple background indicates that, although no LLM is used, other AI approach are employed (e.g. NLP for Natural Language Processing). A star (\ding{79}) on green background marks ontologies that are publicly available in online repositories. The \textbf{OWL Type} column specifies the formal ontology language adopted (e.g., OWL Lite, OWL DL, or OWL Full).
The \textbf{Ontology Modeling} indicates whether the ontology models aspects of the domain: (S) static or (D) dynamic.
The \textbf{Semantic Solution} column distinguishes between ontologies that provide only vocabulary semantics (VS) and those that also define an operational system or application (VS+S). The \textbf{Formal Approach} field identifies whether a reasoning framework is employed, such as logical inference (LI), temporal logic (TL), a fuzzing framework (FF), or enhanced reasoning mechanisms based on temporal logic.}
\label{tab:domian_security_management_and_governance}
\end{table}

% 4.6.1 Attack modeling, techniques, and countermeasures
\subsubsection{\textbf{Attack modeling, techniques, and countermeasures}}

%2012 - Ontology-based modeling of DDoS attacks for attack plan detection
%An ontology-based method to model and proactively detect Distributed Denial of Service (DDoS) attacks is presented by Ansarinia et al. \cite{ansarinia2012ontology}. The approach uses semantic modeling to describe attack patterns, vulnerabilities, and weaknesses. It includes a ruele-based parser that, integrates established taxonomies such as CAPEC, CWE, and CVE into a unified knowledge base. A rule-based parser automatically constructs the ontology by linking relevant concepts. Sensory data, like event logs, is converted into RDF triples for semantic reasoning. It also includes detection rules, written in SWRL that define conditions for identifying specific attack plans. The system reasons over the knowledge base to infer potential attack scenarios. Unlike ML, this method offers interpretability and does not require extensive training. Evaluation with OntoQA metrics confirms its structural richness and semantic depth. The approach supports real-time detection, attack plan recognition, and proactive mitigation.
Ansarinia et al. \cite{ansarinia2012ontology} present an ontology-based method to model and proactively detect Distributed Denial of Service (DDoS) attacks. Their approach semantically models attack patterns, vulnerabilities, and weaknesses using a rule-based parser. Taxonomies such as CAPEC, CWE, and CVE are integrated into a unified knowledge base for reasoning. Sensory data like event logs are converted into RDF triples, enabling semantic detection via SWRL rules. The system infers potential attack scenarios with interpretability, unlike ML approaches that require training.

%2016 - TAON: An ontology-based approach to mitigating targeted attacks
TAON is an ontology-based model for mitigating Advanced Persistent Threats (APTs), covering the full attack lifecycle across actions, assets, and organizational context \cite{luh2016taon}. Its novelty lies in goal modeling, which aligns attacker behavior with strategic objectives, and in leveraging reasoning engines like FaCT++ \cite{factplusplus} for automated threat classification. By mapping events and anomalies to data sources, TAON enhances situational awareness and supports real-world analyst queries to link threats with assets and monitoring tools. Compared to Ansarinia’s SWRL-based DDoS framework, TAON offers a holistic, multi-stage, and dynamic representation that advances ontology-driven threat analysis.

%2022 - Building cybersecurity ontology for understanding and reasoning adversary tactics and techniques
A novel approach to ontology population from unstructured CTI reports is introduced by Huang et al. \cite{huang2022building}. Unlike Ansarinia et al., who rely on predefined rules for DDoS detection, Huang et al. use NLP techniques for knowledge extraction. Their framework surpasses Luh et al.’s manually structured TAON by integrating real-world data through coreference resolution and triplet extraction. It builds on the MITRE ATT\&CK framework, modeling Tactics, Techniques, Software, Groups, and Mitigations. A key novelty is the automated mapping of threat intelligence to formal ontological concepts. The system includes a web platform for interactive querying and threat exploration. It links adversary behaviors, malware, and threat actors more dynamically than prior models. 

%2023 - The Design of an Ontology for ATT&CK and its Application to Cybersecurity
Akbar et al. \cite{akbar2023design} present an ontology that integrates the MITRE ATT\&CK framework with CVE data to support cyber threat response. The key innovation is mapping tactics, techniques, and sub-techniques directly to CVEs for actionable mitigation. Unlike prior works on attack modeling or APT lifecycles, this approach emphasizes vulnerability linkage. Manual CVE integration is currently used, with plans for NLP-based automation inspired by Huang’s CTI extraction. The ontology helps analysts prioritize patching based on attacker behavior and known system flaws. Unlike Luh’s behavioral focus, Akbar centers on real-time, vulnerability-driven decision support. 

% 4.6.2 Risk management, including modeling, assessment, analysis and mitigation
\subsubsection{\textbf{Risk management, including modeling, assessment, analysis and mitigation}}

%2018 - Common Ontology of Value and Risk
A well-founded ontology that formalizes core assumptions about value and risk is introduced by Sales et al. \cite{sales2018common}. They define three perspectives: experiential (events and causes), relational (subjectivity), and quantitative (measurable scales). This ontology clarifies the conceptual foundations of value-risk modeling across different contexts. It offers real-world semantics for established modeling languages like CORAS \cite{lund2010model}, RiskML \cite{siena2014modelling}, the Goal-Risk Framework \cite{asnar2011goal}, and Archimate \cite{band2015modeling}. Beyond semantics, the ontology serves as a reference model for evaluating and redesigning risk modeling approaches. Following established methods, it supports assessments of domain relevance and user comprehensibility. This contributes to improving conceptual clarity and consistency in risk representation. 

%2019 - Ontology-based system for dynamic risk management in administrative domains
A dynamic, ontology-based framework for real-time cybersecurity risk management is proposed by Vega et al. \cite{vega2019ontology}. Unlike Sales et al., who focus on conceptual modeling of value and risk, it emphasizes practical implementation. Their system integrates OWL ontologies with live data feeds (e.g., Syslog \cite{abe2019hayabusa}, Common Event Format) to model diverse assets like IoT devices and citizens. It uses probabilistic inference to link threats to assets when direct associations are missing. Over 2,500 SWRL rules and a decay-based memory function enable reasoning and trend smoothing over time. This work bridges semantic modeling with operational risk analysis. Its key contribution lies in integrating reasoning, uncertainty, and real-time performance in cybersecurity contexts.

%2020 - An ontology-based security risk management model for information systems
CPARR is an ontology-based model that integrates Case-Based Reasoning (CBR) to enable adaptive cybersecurity risk management \cite{arogundade2020ontology}. It classifies IDS-derived threat data using a semantic knowledge base aligned with CVE and CAPEC, computing risk levels via likelihood–impact matrices and frequency analysis.
CBR recommends countermeasures based on past incidents, while an agent-based architecture automates processes and integrates with SIEM systems for scalable deployment. Validated in an e-banking simulation, CPARR demonstrates proactive, context-aware defense that complements Sales’ conceptual rigor and Vega’s real-time system with adaptive, experiential reasoning.

%2022 - An ontology of security from a risk treatment perspective
%ROSE is a formal ontology for security engineering that emphasizes risk treatment and control mechanisms \cite{oliveira2022ontology}. Unlike Sales et al.'s broad conceptualization of value and risk, ROSE focuses on how security mechanisms prevent or mitigate threats. Built on UFO \cite{guizzardi2022ufo} and COVER \cite{Sales2018COVR}, it enhances semantic clarity, generality, and alignment with FAIR principles. A key innovation is modeling security mechanisms as control-capable objects, capturing their role in reducing harmful events. ROSE distinguishes between intentional and unintentional threats, aiding in precise countermeasure selection. It supports ISO 31000\footnote{https://www.iso.org/standard/65694.html}
%risk strategies and models residual risk for iterative security evaluation. By linking threats, vulnerabilities, controls, losses, and agent goals, ROSE facilitates probabilistic risk assessment.
ROSE is a formal ontology for security engineering that emphasizes risk treatment and control mechanisms \cite{oliveira2022ontology}. Unlike Sales et al.’s broad value–risk conceptualization, it focuses on how security mechanisms prevent or mitigate threats.
Built on UFO \cite{guizzardi2022ufo} and COVER \cite{Sales2018COVR}, ROSE enhances semantic clarity, generality, and FAIR alignment. Its key innovation is modeling security mechanisms as control-capable objects that reduce harmful events. ROSE distinguishes intentional from unintentional threats, supporting ISO 31000\footnote{https://www.iso.org/standard/65694.html}
 strategies and residual risk modeling. By linking threats, vulnerabilities, controls, losses, and goals, it enables probabilistic risk assessment.

%2023 - Ontology-based approach to real-time risk management and cyber-situational awareness
Sánchez-Zas et al. \cite{sanchez2023ontology} propose an ontology-based framework for real-time risk management and cyber situational awareness. Building on the work of Vega et al., they replace SWRL with SPIN\footnote{https://www.w3.org/submissions/spin-overview/} rules to enhance reasoning performance in time-sensitive contexts. The framework incorporates dynamic anomaly detection through sensor ontologies and unifies physical and logical threat analysis to support comprehensive risk inference in heterogeneous systems. A key contribution is the integration of anomaly detection with SPIN-based reasoning, enabling real-time threat response. The approach supports both residual and potential risk assessments, aligning with best practices in situational awareness and the operationalization of threat intelligence.

%2025 - Toward an ontology-based modeling for risk management
The most recent work of Oliveira et al. \cite{oliveira2025toward} present a comprehensive ontology network to overcome the semantic limitations of traditional techniques like attack trees and Failure Mode and Effects Analysis (FMEA). Grounded in UFO, it unifies domains such as trust, resilience, and incidents, extending beyond the risk treatment scope of ROSE.
Unlike Vega et al. and Sánchez-Zas et al., who prioritize real-time systems, this work emphasizes semantic precision, formal analysis, and interoperability. A domain-specific modeling language enables formal representation of hypothetical and real incidents, supporting reasoning, simulation, and root cause analysis with CVE, CAPEC, and ATT\&CK integration.
Its key contribution lies in unifying diverse domains through an ontology network that bridges foundational modeling with operational cybersecurity needs.

% 4.6.3 Threats and vulnerability modeling
\subsubsection{\textbf{Threats and vulnerability modeling}}

%2019 - Threat modeling the cloud: An ontology based approach
An ontology-based threat modeling approach for cloud environments, with emphasis on critical infrastructure, is presented by Manzoor et al. \cite{manzoor2019threat}. The method captures relationships among users, cloud services, and threats, mapping vulnerabilities to confidentiality, integrity, and availability objectives. User requirements are prioritized by criticality and linked to cloud services modeled in OpenStack, such as virtual machine provisioning. By mapping the ontology to a Design Structure Matrix (DSM), the framework visualizes dependencies, supports multi-perspective analysis, and highlights influential actors through threat propagation.

%2020 - Automating threat modeling using an ontology framework
Valja et al. \cite{valja2020automating} present an ontology-based framework for automating threat modeling and improving cybersecurity data quality.
Unlike Manzoor et al., who rely on manually built ontologies and DSMs, this work offers a fully automated pipeline addressing semantic inconsistencies, data granularity, and heterogeneous sources. The framework integrates ontology patterns, reasoning, and data enrichment, using ArangoDB and Python-based logic to classify OSs, applications, vulnerabilities, and data flows.
Validated on industrial and academic datasets, it improves precision, reduces manual effort, and integrates with SecuriCAD\footnote{https://www.foreseeti.com/securicad/}
 for automated risk analysis. This approach advances scalable, ontology-driven cybersecurity modeling by combining automation, coherence, and practical applicability.

%2022 - Threma: Ontology-based automated threat modeling for ict infrastructures
ThreMA, an ontology-based framework for dynamic, real-time threat modeling in complex ICT infrastructures, is introduced by Rosa et al. \cite{de2022threma}. It enhances scalability and automation through semantic integration and reasoning, unifying data from vulnerability databases, network topologies, and asset inventories. A formal ontology with semantic rules maps vulnerabilities to infrastructure components and infers missing information. The framework automates data ingestion, threat graph generation, and visualization, enabling adaptive and proactive threat identification. 

%2024 - Ontology-driven Threat Modeling for IoT Systems
An ontology-based framework for automated threat modeling in IoT systems is proposed by Compierchio \cite{compierchio2024ontology}.
It addresses IoT-specific challenges by modeling heterogeneous devices, limited resources, and diverse communication paths with OWL 2 ontologies.
The framework comprises three sub-ontologies: IoT System (aligned with ISO/IEC 30141\footnote{https://www.iso.org/standard/88800.html}
), Data Flow, and Threats (mapping STRIDE \cite{abuabed2023stride} to CAPEC).
SWRL rules support automated reasoning to infer threats from configurations and interactions, validated on the HArMoNICS smart microgrid.
The system auto-generates context-aware threat models, offering a reusable and extensible method that enhances automation and semantic precision in IoT security.

%Conclusion
\paragraph{\textbf{Conclusion for the Security Management and Governance domain.}} In the Security Management and Governance domain, ontologies are universally adopted, with varying degrees of advancement across subdomains. In attack modeling, techniques, and countermeasures, contributions rely on static DL ontologies that support VS+S solutions, with one notable case integrating NLP/LLM methods, marking an early attempt at hybridization. In risk management, including modeling, assessment, analysis, and mitigation, the approaches are more diverse, with both static and dynamic ontologies. Several studies adopt TL to enhance reasoning, and one work explicitly reuses a publicly available ontology, which is rare across domains. Some contributions also integrate KGs, showing more openness to linked and evolving data models. This subdomain demonstrates the highest methodological rigor within the domain. For threats and vulnerability modeling, the emphasis remains on static ontologies with VS+S implementations that provide practical tools and applications. However, no formal reasoning frameworks, reuse, or hybrid AI methods are present here.

\definecolor{lightorange}{RGB}{255,245,230}
\begin{tcolorbox}[
  colback=lightorange,   % background color
  colframe=lightorange,  % border color
  boxrule=0pt,           % no visible border
  arc=0mm,               % sharp corners
  left=2mm, right=2mm,   % padding
  top=1mm, bottom=1mm
]
\textbf{Summary:} Table \ref{tab:domian_security_management_and_governance} shows that this domain maintains a steady reliance on ontologies with strong application orientation (VS+S), complemented in some subdomains by early adoption of formal reasoning TL, KG integration, and NLP/LLM hybridization. Progress, however, is uneven: risk management emerges as the most advanced area, while threat and vulnerability modeling remain comparatively static. The absence of systematic ontology reuse, semantic log integration, and large-scale hybrid symbolic–statistical approaches continues to limit the domain’s potential for adaptive and real-time governance support.
\end{tcolorbox}

% 7 Software and Hardware Security Engineering 
\subsection{Software and Hardware Security Engineering}

This domain covers security throughout the entire lifecycle of software and hardware systems, ranging from risk assessment and requirements analysis to secure architecture design, implementation, validation, and testing. It also includes deployment and runtime monitoring to ensure continued protection during system operation \cite{JRC2021CybersecurityTaxonomy, Nai2019}. The goal is to build resilient, secure-by-design systems that uphold confidentiality, availability, and integrity at every stage. The domain encompasses 20 subdomains, of which our work compiles existing studies and ontologies for 3, based on the criteria defined in Section \ref{sec:2.2} and summarized in:
\begin{inparaenum}[(\bgroup\bfseries 1\egroup)]
\item \emph{Attack techniques}, which represent collection of methods used by adversaries to exploit vulnerabilities, including technical exploits such as buffer overflows, code injection, privilege escalation, and remote code execution, as well as social engineering tactics like phishing and spam to deceive users and gain unauthorized access;
\item \emph{Intrusion detection and honeypots} consists of technique and systems used to detect unauthorized access or malicious activity (intrusion detection) and to lure, observe, and analyze attacker behavior using decoy systems (honeypots); and
\item \emph{Malware analysis including adversarial learning of malware} is the process of examining malicious software to understand its structure, behavior, and impact, including techniques to detect, classify, and mitigate threats. It also encompasses adversarial learning, where ML models are trained or attacked using crafted malware variants to improve detection robustness or evade security systems.
\end{inparaenum}
Table~\ref{tab:software_and_hardware_security_engineering} provides a comparative summary of research in the Security Management and Governance domain, highlighting the inclusion and characteristics of core elements such as ontologies, KGs, semantic reasoning, and formal methods.

% Table
\newcolumntype{C}[1]{>{\centering\arraybackslash}p{#1}}

\begin{table}[h]
\centering
\rowcolors{2}{gray!10}{white}
\begin{tabular}{p{3.8cm} C{1.1cm} C{0.8cm} C{1.cm} C{1cm} C{1.4cm} C{1.4cm} C{1.5cm} C{1.4cm}}
\toprule
\textbf{Domain, Subdomain and Work} & \textbf{Ontology} & \textbf{KG} & \textbf{Semantic Log} & \textbf{LLM} & \textbf{OWL Type} & \textbf{Ontology Modeling} & \textbf{Semantic Solution} & \textbf{Formal Approach} \\
\midrule
\rowcolor{gray!30}\multicolumn{9}{l}{ \textbf{Software and Hardware Security Engineering}} \\

%Attack techniques
\midrule
\rowcolor{lightgray}\multicolumn{9}{l}{ \textbf{Attack techniques}} \\
\midrule
%Spam
Youn \cite{youn2014spongy} & \cellcolor{green!25}\ding{51} & \cellcolor{red!25}\ding{53} & \cellcolor{green!25}\ding{51} & \cellcolor{red!25}\ding{53} & -- & D & VS+S & FF \\

Agrawal et al. \cite{agrawal2022ontospammer} & \cellcolor{green!25}\ding{51} & \cellcolor{red!25}\ding{53} & \cellcolor{green!25}\ding{51} & \cellcolor{green!25}\ding{51}  & -- & D & VS+S & FF \\

Venvckauskas et al. \cite{venvckauskas2024email} & \cellcolor{green!25}\ding{51} & \cellcolor{red!25}\ding{53} & \cellcolor{green!25}\ding{51} & \cellcolor{blue!25}\ding{53} ML & -- & D & VS+S & FF \\

%Phishing
Mahdi et al. \cite{bazarganigilani2011phishing} & \cellcolor{green!25}\ding{51} & \cellcolor{red!25}\ding{53} & \cellcolor{green!25}\ding{51} & \cellcolor{blue!25}\ding{53} ML & -- & S & VS & -- \\

Park et al. \cite{park2018ontological} & \cellcolor{green!25}\ding{51} & \cellcolor{red!25}\ding{53} & \cellcolor{green!25}\ding{51} & \cellcolor{blue!25}\ding{53} NaB & -- & S & VS & -- \\

Tchakounté et al. \cite{tchakounte2020description} & \cellcolor{green!25}\ding{51} & \cellcolor{red!25}\ding{53} & \cellcolor{red!25}\ding{53} & \cellcolor{red!25}\ding{53} & DL & S & VS+S & -- \\

Oliveira et al. \cite{oliveira2023toward} & \cellcolor{green!60}\ding{79} \cite{utwente_phishing_ontology_2025} & \cellcolor{red!25}\ding{53} & \cellcolor{red!25}\ding{53} & \cellcolor{red!25}\ding{53} & -- & S & VS+S & -- \\

Zahedi et al. \cite{zahedi2024ontology} & \cellcolor{green!25}\ding{51} & \cellcolor{red!25}\ding{53} & \cellcolor{red!25}\ding{53} & \cellcolor{red!25}\ding{53} & -- & S & VS+S & -- \\

% Intrusion detection and honeypots
\midrule
\rowcolor{lightgray} \multicolumn{9}{l}{\textbf{Intrusion detection and honeypots}} \\
\midrule

Abdoli et al. \cite{abdoli2009ontology} & \cellcolor{green!25}\ding{51} & \cellcolor{red!25}\ding{53} & \cellcolor{red!25}\ding{53} & \cellcolor{red!25}\ding{53} & -- & S & VS+S & -- \\

Tailhardat et al. \cite{tailhardat2024noria} & \cellcolor{green!25}\ding{51} & \cellcolor{green!25}\ding{51} & \cellcolor{green!25}\ding{51} & \cellcolor{blue!25}\ding{53} NLP & DL & D & VS+S & -- \\

Ayo et al. \cite{ayo2024ontology} & \cellcolor{green!25}\ding{51} & \cellcolor{red!25}\ding{53} & \cellcolor{green!25}\ding{51} & \cellcolor{red!25}\ding{53} & -- & D & VS+S & -- \\

%Sikos et al. \cite{sikos2019knowledge} & \cellcolor{green!60}\ding{79} \cite{PAO_ontology} &  \cellcolor{green!25}\ding{51} &  \cellcolor{green!25}\ding{51} & \cellcolor{red!25}\ding{53} & DL & S & VS & -- \\

Andrew et al. \cite{andrew2023knowledge} & \cellcolor{green!25} &  \cellcolor{green!25}\ding{51} &  \cellcolor{green!25}\ding{51} & \cellcolor{red!25}\ding{53} & -- & S & VS+S & -- \\

% Malware analysis including adversarial learning of malware
\midrule
\rowcolor{lightgray}\multicolumn{9}{l}{ \textbf{Malware analysis including adversarial learning of malware}} \\
\midrule

Mundie et al. \cite{mundie2013ontology} & \cellcolor{green!25}\ding{51} & \cellcolor{red!25}\ding{53} & \cellcolor{green!25}\ding{51} & \cellcolor{red!25}\ding{53} & DL & S & VS & -- \\

Gregio et al. \cite{gregio2014ontology} & \cellcolor{green!25}\ding{51} & \cellcolor{red!25}\ding{53} & \cellcolor{red!25}\ding{53} & \cellcolor{red!25}\ding{53} & -- & S & VS+S & -- \\

Rastogi et al. \cite{rastogi2020malont} & \cellcolor{green!25}\ding{51} & \cellcolor{green!25}\ding{51} & \cellcolor{red!25}\ding{53} & \cellcolor{red!25}\ding{53} & DL & D & VS+S & -- \\

\midrule
\end{tabular}
\caption{Overview of research efforts in the \textbf{Software and Hardware Security Engineering} domain.
A checkmark (\ding{51}) on a green background indicates the presence of a component (e.g., \textbf{Ontology}, \textbf{Knowledge Graph}, \textbf{Semantic Log}, or \textbf{LLM}). A cross (\ding{53}) on a red background denotes its absence, while a cross on a purple background indicates that, although no LLM is used, other AI approach are employed (e.g. ML for Machine Learning, NaB for Naive Bayes and DL for Deep Learning). A star (\ding{79}) on green background marks ontologies that are publicly available in online repositories. The \textbf{OWL Type} column specifies the formal ontology language adopted (e.g., OWL Lite, OWL DL, or OWL Full).
The \textbf{Ontology Modeling} indicates whether the ontology models aspects of the domain: (S) static or (D) dynamic.
The \textbf{Semantic Solution} column distinguishes between ontologies that provide only vocabulary semantics (VS) and those that also define an operational system or application (VS+S). The \textbf{Formal Approach} field identifies whether a reasoning framework is employed, such as logical inference (LI), temporal logic (TL), a fuzzing framework (FF), or enhanced reasoning mechanisms based on temporal logic.}
\label{tab:software_and_hardware_security_engineering}
\end{table}

% 4.7.1 Attack techniques
\subsubsection{\textbf{Attack techniques}}

% 2020 Defining social engineering in cybersecurity
In cybersecurity, \emph{social engineering} refers to an attack technique in which adversaries manipulate human behavior to exploit vulnerabilities and compromise fundamental security objectives such as confidentiality, integrity and availability \cite{wang2020defining}. Within the attack techniques subdomain, we highlight several studies that explore spam and phishing attacks in the broader context of social engineering and related threat vectors.

%2014 - SPONGY (SPam ONtoloGY): Email Classification Using Two-Level Dynamic Ontology}
SPONGY (SPam ONtoloGY), a two-tier dynamic ontology-based approach for email spam filtering, is introduced by Youn \cite{youn2014spongy}. The system integrates a global ontology with user-specific ontology filters to improve classification accuracy. The paper outlines the system architecture, data processing workflow, and ontology construction methods, leveraging the \emph{Weka}\footnote{https://ml.cms.waikato.ac.nz/weka/} toolkit and the \emph{Jena}\footnote{https://jena.apache.org/} framework.

%ML
%2022 - OntoSpammer: A Two-Source Ontology-Based Spam Detection Using Bagging
OntoSpammer is a spam detection system that combines a dual ontology model with ensemble learning, proposed by Agrawal et al. \cite{agrawal2022ontospammer}. It integrates domain and linguistic ontologies to enrich semantic features and address the shortcomings of prior models. Unlike SPONGY, which relied on global and user-specific filters, OntoSpammer adopts a two-source ontology design. Robustness is improved through bagging with classifiers such as SVM \cite{noble2006support} and Decision Trees \cite{de2013decision}. Moving beyond traditional tools like \emph{Weka}, it leverages advanced ensemble techniques for semantic and model diversity.

%2024 - An Email Cyber Threat Intelligence Method Using Domain Ontology and Machine Learning
Venvckauskas et al. \cite{venvckauskas2024email} propose a domain-specific ontology and a privacy-preserving method for spam detection. Unlike Agrawal et al., their approach relies on email metadata rather than message content. This makes it suitable for encrypted or privacy-sensitive scenarios.
They develop a semantic parser for language-independent classification using header metadata. In contrast to SPONGY and OntoSpammer’s content-based methods, this avoids reliance on NLP. The method enhances Cyber Threat Intelligence (CTI) sharing without exposing private content. By focusing on metadata, it overcomes limitations of text-dependent ontological models. This marks a key shift toward privacy-aware, ontology-based spam detection.

% Phishing
Additionally, \emph{Phishing} the most prevalent forms of cyberattacks, raising growing concerns as many internet users fall victim to such schemes \cite{chiew2018survey}.

%2011 - Phishing E-Mail Detection Using Ontology Concept and Naïve Bayes Algorithm
%Text Mining approach
Mahdi et al. \cite{bazarganigilani2011phishing} propose a phishing email detection method combining ontology-based semantic enrichment with the Naïve Bayes algorithm \cite{murphy2006naive}. Emails are first screened for common phishing indicators such as hex-based IP URLs or mismatched sender details.
Unclassified emails are then analyzed through text classification enhanced with ontology concepts. Ontologies map words to higher-level concepts, mitigating synonymy and ambiguity. A context-based disambiguation strategy selects the most relevant meaning of each word. Finally, Term Frequency Variance (TFV) reduces dimensionality, and the Naïve Bayes classifier performs the final classification.

%2018 - Ontological Detection of Phishing Emails
An ontology-based method for phishing email detection focused on semantic understanding is presented by Park et al. \cite{park2018ontological}. Their approach identifies high-impact verbs using Term Frequency-Inverse Document Frequency (TF-IDF) \cite{qaiser2018text} and analyzes argument structures to infer intent. Stanford CoreNLP \cite{song2014text} is used for syntactic parsing and dependency extraction to support deep semantic analysis. Unlike Mahdi et al., who apply ontology-enhanced Naïve Bayes on term features, Park et al. captures verb-driven semantic patterns within an ontological framework. This enables accurate interpretation of actions like "verify" or "update" to detect phishing behavior. The method enhances robustness against keyword manipulation and improves interpretability.

%2020 - A description logic ontology for email phishing
A formal ontology for phishing detection is proposed by Tchakounté et al. \cite{tchakounte2020description}. Unlike Mahdi’s statistical model and Park’s NLP-driven approach, this framework emphasizes logical reasoning and structured knowledge. Namelly DL, it is used to model phishing behaviors through defined concept hierarchies and semantic relationships. The ontology supports automated inference, consistency checking, and knowledge base expansion. This logic-based design improves interpretability and enables machine-actionable intelligence.
It moves beyond keyword patterns toward a rule-governed, context-aware detection model. The framework provides a robust foundation for adaptable and explainable phishing detection.

%2023 - Toward a phishing Attack Ontology
Oliveira et al. \cite{oliveira2023toward} introduce PHATO\footnote{https://github.com/utwente-scs/phishing-ontology}, a phishing ontology addressing structural and semantic gaps in earlier models. It integrates with established ontologies like ROSE\footnote{https://github.com/unibz-core/security-ontology} and COVER\footnote{https://github.com/unibz-core/value-and-risk-ontology}. PHATO supports both the representation of phishing attacks and design of countermeasures. Its conceptual precision enables broad applicability beyond detection tasks.

%2024 - Ontology-Based Intelligent Interface Personalization for Protection Against Phishing Attacks
The Ontology-based Intelligent Interface Personalization (OBIIP) framework, designed to improve phishing detection tool usability through intelligent interface personalization, is introduced by Zahedi et al. \cite{zahedi2024ontology}. Unlike Park and Tchakounté, who focus on detection models, this work emphasizes human-centered security and user trust. The authors develop the Ontology of Warning Interface Elements (OWIE) based on empirical data. OWIE guides the design of adaptive interfaces that support informed user decision-making. A two-phase evaluation shows improvements in user engagement, trust, and self-protection. The framework bridges ontology engineering with Human-Computer Interaction (HCI). It addresses a critical gap in user experience within phishing defense systems.

% 4.7.2 Intrusion detection and honeypots
\subsubsection{\textbf{Intrusion detection and honeypots}}

% 2009 - Ontology-based distributed intrusion detection system
A Distributed Intrusion Detection System (DIDS) that leverages ontologies to improve attack detection is presented by Abdoli et al. \cite{abdoli2009ontology}. The system includes multiple IDS agents coordinated by a central MasterAgent with semantic analysis capabilities. It targets Denial of Service (DoS) attacks by leveraging semantic relationships among network events. This enhances detection accuracy while reducing false positives and negatives. Evaluation with the KDD 99 dataset \cite{olusola2010analysis} shows superior performance over traditional methods. The architecture enables intelligent, collaborative decision-making among agents. It supports dynamic ontology updates to address evolving threats.

% 2024 - NORIA-O: An Ontology for Anomaly Detection and Incident Management in ICT Systems
% SEAS - \cite{lefranccois2017seas}
% UCO - \cite{syed2016uco}

NORIA-O, a semantic anomaly detection framework that diverges from purely data-driven approaches, is presented by Tailhardat et al. \cite{tailhardat2024noria}. It introduces an ontology-based model that represents infrastructure, incidents, and organizational context. By integrating multiple domain ontologies (e.g., SEAS\footnote{https://ci.mines-stetienne.fr/seas/}, UCO\footnote{https://github.com/ucoProject/UCO}, ORG\footnote{https://www.w3.org/TR/vocab-org/}, BBO\footnote{https://raw.githubusercontent.com/ProfTuan/BBO/refs/heads/main/BBO.owl}), NORIA-O enables structural reasoning beyond syntactic or temporal log patterns. The framework supports formal inference about the causes, impacts, and relationships of anomalies within the system architecture. It provides a modular, semantic foundation for incident analysis, emphasizing root cause diagnosis, impact assessment, and coordinated mitigation. Through ontological event embedding, NORIA-O enhances interpretability, contextual awareness, and decision support.

% 2024 - Ontology-Based Layered Rule-Based Network Intrusion Detection System for Cybercrimes Detection
%ML
%Ayo et al. \cite{ayo2024ontology} propose a hybrid ontology-based NIDS combining semantic modeling with machine learning.
%Building on Abdoli et al.’s static, rule-based approach, they enable dynamic rule generation via genetic algorithms, neural networks, and decision trees.
%The system, trained on the UNSW-NB15  dataset\footnote{https://research.unsw.edu.au/projects/unsw-nb15-dataset}
%, automates feature selection and rule creation. These rules are semantically formalized and embedded within a modular ontology.
%This supports context-aware, machine-interpretable detection of cybercrime.
%The ontology enhances scalability, adaptability, and structured reasoning over attack behaviors. It enables automatic, up-to-date rule generation and formal representation of attack concepts.
%Unlike A2Log and MLog’s fully data-driven log analysis, Ayo et al. integrate statistical learning with symbolic inference. Their framework bridges ML and ontological reasoning, enhancing interpretability and reuse. It provides an extensible solution for network-based anomaly detection in dynamic threat contexts. The ontology-driven structure adapts with new attack signatures, concepts, and relationships without full re-engineering.

Ayo et al. \cite{ayo2024ontology} propose a hybrid ontology-based NIDS that combines semantic modeling with machine learning. Unlike Abdoli et al.’s static rules, they enable dynamic rule generation via genetic algorithms, neural networks, and decision trees.
Trained on the UNSW-NB15 dataset\footnote{https://research.unsw.edu.au/projects/unsw-nb15-dataset}, the system automates feature selection and semantic rule creation.
These rules are embedded in a modular ontology, supporting context-aware, machine-interpretable detection of cybercrime. The framework bridges ML with symbolic inference, enhancing interpretability, scalability, and structured reasoning. It adapts with new attack signatures and concepts, offering an extensible solution for dynamic threat contexts.

A framework that applies knowledge graphs to analyze data from honeypots such as Cowrie\footnote{https://github.com/cowrie/cowrie} and Dionaea\footnote{https://dionaea.readthedocs.io/en/latest/installation.html} is presented by Andrew et al. \cite{andrew2023knowledge}. Honeypots generate rich logs of attacker activity, which are difficult to interpret using traditional methods. The framework structures this data into a semantic graph enriched with ontologies, capturing entities such as IPs, commands, sessions, and honeypot types.
Using Neo4j\footnote{https://neo4j.com/}, analysts can query command sequences, summarize attacker behavior, and map attack origins. Ontology-based reasoning enhances detection of patterns, anomalies, and potential attribution of attacks. The study concludes that integrating honeypots with knowledge graphs and ontologies strengthens cybersecurity analysis.

% 4.7.3 Malware analysis including adversarial learning of malware
\subsubsection{\textbf{Malware analysis including adversarial learning of malware}}

%2013 - An Ontology for Malware Analysis
The lack of a shared vocabulary in cybersecurity, particularly in malware analysis, is addressed by Mundie et al. \cite{mundie2013ontology}. They propose malware analysis ontology to standardize terminology and structure domain knowledge. They define six levels of knowledge in a general context as a methodological framework for knowledge representation. The current ontology on malware analysis incorporates four of these: controlled vocabulary, taxonomy, static ontology (for competency modeling), and intentional ontology. This layered approach supports both descriptive and inferential reasoning in malware analysis tasks. It lays the groundwork for enhanced interoperability, consistency, and automation in cyber threat understanding.

%2014 - Ontology for Malware Behavior: a Core Model Proposal
Ontology-based malware analysis is advanced by Gregio et al. \cite{gregio2014ontology} through dynamic behavioral modeling. They introduce a core behavioral ontology capturing actions like evasion, stealing, and subversion. This approach moves beyond rigid taxonomies, enabling nuanced reasoning and flexible analysis. It enhances real-time detection and incident response through semantically rich behavior traces.
A practical use case demonstrates its operational value beyond theoretical design.

%2020 - MALOnt: An Ontology for Malware Threat Intelligence
Malware ontologies are advanced by integrating real-world threat intelligence, as shown by Rastogi et al. \cite{rastogi2020malont}.
Building on prior work like Mundie et al. (domain structuring) and Gregio et al. (behavior modeling), they propose MALOnt. MALOnt offers an open-source, extensible architecture for analyzing heterogeneous malware data.
It uniquely supports instance-level enrichment, such as attacker profiles, IOCs, and malware samples. This allows the creation of dynamic knowledge graphs for reasoning and machine learning. By bridging symbolic and statistical AI, MALOnt enhances threat correlation and prediction.
Its automated reasoning enables real-time threat detection and analysis, shifting towards intelligence-driven frameworks.

%Conclusion
\paragraph{\textbf{Conclusion for the Software and Hardware Security Engineering domain.}} In this domain, ontologies are consistently adopted across all subdomains, but integration with KGs, semantic logs, and hybrid AI methods is uneven. In the attack techniques subdomain, most contributions use dynamic ontologies and deliver VS+S solutions, with several works experimenting with ML integration, NLP, or Naïve Bayes (NaB) approaches. Some also employ FF frameworks, and one ontology is explicitly publicly available, showing progress in openness and methodological rigor. In the subdomain of intrusion detection and honeypots, contributions rely mainly on static ontologies and VS+S implementations, with one study adopting NLP integration. Ontology reuse appears here, but overall reasoning and formal approaches are absent. For malware analysis, including adversarial learning of malware, studies remain more conservative, focusing on static ontologies. Most deliver VS+S solutions, while some limit their work to vocabulary semantics (VS). ML/LLM integration, semantic logs, and formal reasoning are not yet explored in this area.

%Summary
\definecolor{lightorange}{RGB}{255,245,230}
\begin{tcolorbox}[
  colback=lightorange,   % background color
  colframe=lightorange,  % border color
  boxrule=0pt,           % no visible border
  arc=0mm,               % sharp corners
  left=2mm, right=2mm,   % padding
  top=1mm, bottom=1mm
]
\textbf{Summary:} Table \ref{tab:software_and_hardware_security_engineering} shows a domain with a strong reliance on ontologies with applied VS+S solutions and shows some innovation through hybridization with ML, NLP, and formal FF, particularly in the attack techniques subdomain. However, the lack of widespread adoption of KGs, semantic logs, advanced reasoning frameworks, and systematic ontology reuse constrains broader interoperability and the capacity to support adaptive, real-time security engineering tasks.
\end{tcolorbox}

% 8 - Theoretical Foundations 
\subsection{Theoretical Foundations}
This domain focuses on the application of formal methods for the analysis and verification of security properties. It aims to provide mathematical guarantees of correctness in software, hardware, and algorithms \cite{JRC2021CybersecurityTaxonomy, Nai2019}. Such techniques help identify vulnerabilities and ensure compliance with security specifications. They are essential for building provably secure systems and protocols. It encompasses 6 subdomains, within which this work aggregates prior studies and ontologies for 2, based on the criteria defined in Section \ref{sec:2.2} and described:
\begin{inparaenum}[(\bgroup\bfseries 1\egroup)]
\item \emph{Formal specification, analysis, and verification of software and hardware} involve using mathematical and logical methods to precisely define system behavior, rigorously analyze design and implementation, and formally prove that software and hardware meet intended functionality and security properties—such as confidentiality, integrity and availability—under all conditions; and
\item \emph{Formal verification of security assurance} is the use of mathematical proofs and formal methods to rigorously demonstrate that a system’s security mechanisms meet defined assurance requirements, ensuring that security objectives (e.g., access control, data protection, isolation) are correctly and consistently implemented.
\end{inparaenum}

% Table
\newcolumntype{C}[1]{>{\centering\arraybackslash}p{#1}}

\begin{table}[h]
\centering
\rowcolors{2}{gray!10}{white}
\begin{tabular}{p{3.8cm} C{1cm} C{0.8cm} C{1cm} C{1cm} C{1.4cm} C{1.4cm} C{1.5cm} C{1.4cm}}
\toprule
\textbf{Domain, Subdomain and Work} & \textbf{Ontology} & \textbf{KG} & \textbf{Semantic Log} & \textbf{LLM} & \textbf{OWL Type} & \textbf{Ontology Modeling} & \textbf{Semantic Solution} & \textbf{Formal Approach} \\
\midrule
\rowcolor{gray!30} \multicolumn{9}{l}{\textbf{Theoretical Foundations}} \\

%Attack techniques
\midrule
\rowcolor{lightgray}\multicolumn{9}{l}{ \textbf{Formal specification, analysis, and verification of software and hardware}} \\
\midrule

Tsoukalas et al.~\cite{tsoukalas2021ontology} & \cellcolor{green!25}\ding{51} & \cellcolor{red!25}\ding{53} & \cellcolor{red!25}\ding{53} & \cellcolor{red!25}\ding{53} & DL & S & VS+S & FF \\

Neupane et. al~\cite{neupane2022ontology} & \cellcolor{green!25}\ding{51} & \cellcolor{red!25}\ding{53} & \cellcolor{green!25}\ding{51} & \cellcolor{green!25}\ding{51}  & DL & D & VS+S & TL \\

Brucker et al. \cite{brucker2023using} & \cellcolor{green!25}\ding{51} & \cellcolor{red!25}\ding{53} & \cellcolor{green!25}\ding{51} & \cellcolor{green!25}\ding{51}  & DL & S & VS & -- \\

Mokos et al.~\cite{mokos2025model} & \cellcolor{green!25}\ding{51} & \cellcolor{red!25}\ding{53} & \cellcolor{red!25}\ding{53} & \cellcolor{red!25}\ding{53}  & DL & S & VS+S & -- \\

\midrule
\end{tabular}
\caption{Overview of research efforts in the \textbf{Theoretical Foundations} domain.
A checkmark (\ding{51}) on a green background indicates the presence of a component (e.g., \textbf{Ontology}, \textbf{Knowledge Graph}, \textbf{Semantic Log}, or \textbf{LLM}). A cross (\ding{53}) on a red background denotes its absence, while a cross on a purple background indicates that, although no LLM is used, other AI approach are employed (e.g. ML for Machine Learning and DL for Deep Learning). The \textbf{OWL Type} column specifies the formal ontology language adopted (e.g., OWL Lite, OWL DL, or OWL Full).
The \textbf{Ontology Modeling} indicates whether the ontology models aspects of the domain: (S) static or (D) dynamic.
The \textbf{Semantic Solution} column distinguishes between ontologies that provide only vocabulary semantics (VS) and those that also define an operational system or application (VS+S). The \textbf{Formal Approach} field identifies whether a reasoning framework is employed, such as logical inference (LI), temporal logic (TL), a fuzzing framework (FF), or enhanced reasoning mechanisms based on temporal logic.}
\label{tab:domian_theoretical_foundations}
\end{table}

% 4.8.1 Formal specification, analysis, and verification of software and hardware
\subsubsection{\textbf{Formal specification, analysis, and verification of software and hardware}}

%2021 - An ontology-based approach for automatic specification, verification, and validation of software security requirements: Preliminary results
%An ontology-based approach to address vague or incomplete software security requirements is proposed by Tsoukalas et al. \cite{tsoukalas2021ontology}. The Software Security Requirements Specification (SSRS) uses syntactic and semantic NLP to transform natural language requirements into ontology objects. Its ontology schema captures Actors, Actions, Objects, Properties, Priorities, and Security Characteristics for formal representation. The Software Security Requirements Verification and Validation (SSRVV) then compares these ontology instances against a curated knowledge base. Semantic similarity methods (WordNet\footnote{https://mitpress.mit.edu/9780262561167/wordnet/}, NLTK\footnote{https://www.nltk.org/}) detect inconsistencies, suggest refinements, and recommend priority terms. Both mechanisms are delivered as web services in the IoTAC platform, leveraging AI-driven NLP and enabling future LLM integration.
Tsoukalas et al. \cite{tsoukalas2021ontology} propose an ontology-based approach to handle vague or incomplete software security requirements. The SSRS uses syntactic and semantic NLP to transform natural language requirements into ontology objects.
Its schema models Actors, Actions, Objects, Properties, Priorities, and Security Characteristics for formal representation.
The SSRVV compares these ontology instances against a curated knowledge base for validation.
Semantic similarity methods (WordNet\footnote{https://mitpress.mit.edu/9780262561167/wordnet/}
, NLTK\footnote{https://www.nltk.org/}
) detect inconsistencies and recommend refinements.
Both mechanisms are offered as web services in the IoTAC platform, supporting AI-driven NLP and future LLM integration.

%2022 - An ontology-based framework for formal verification of safety and security properties of control logics
A formal ontology-based verification framework for Industrial Control Systems (ICS) is proposed by Neupane and Mehrpouyan \cite{neupane2022ontology}. Their method integrates OWL-DL and SWRL with model checking to support rich logical representation.
A key innovation is translating SWRL rules into Timed Computational Tree Logic (TCTL) formulas for verification with UPPAAL \cite{david2015uppaal}. This enables semantic-preserving analysis of time-dependent behaviors using Timed Automata.
The Pellet reasoner \cite{parsia2004pellet} ensures consistency, and traceability is maintained across models. The approach improves automation, rigor, and safety assurance in ICS development.

%2023 - Using deep ontologies in formal software engineering
%A novel method that embeds deep ontologies into formal software engineering using Higher-Order Logic (HOL) is introduced by Brucker et al. \cite{brucker2023using}. A deep ontology is an ontology built directly into a formal logic system, allowing knowledge to be both described and automatically checked for consistency. Unlike Tsoukalas et al., who use OWL and NLP for requirements engineering, their approach enables formal reasoning within the Isabelle proof assistant\footnote{https://isabelle.in.tum.de/}. Differing from Neupane and Mehrpouyan’s model-checking, ontologies are directly integrated into the requirements specification through the Ontology Definition Language (ODL). A key contribution is the concept of term-contexts, which enforce semantic constraints within logical terms. This enables machine-checked annotations and document-level semantic validation with strong traceability.
%The method advances native ontology integration for high-assurance and certification-driven software verification.

Brucker et al. \cite{brucker2023using} introduce a method embedding deep ontologies into formal software engineering via Higher-Order Logic (HOL).
A deep ontology is directly built into a formal logic system, enabling knowledge description and consistency checking.
Unlike Tsoukalas et al., who use OWL and NLP, their approach leverages formal reasoning within Isabelle\footnote{https://isabelle.in.tum.de/}. Differing from Neupane and Mehrpouyan, ontologies are integrated into requirements via the Ontology Definition Language (ODL). A key contribution is term-contexts, which enforce semantic constraints and enable machine-checked annotations. This supports document-level semantic validation with strong traceability for high-assurance software verification.

%2025 - Model-based safety analysis of requirement specifications
%A semi-automated framework that links informal requirements to formal safety analysis via ontology-driven model synthesis is proposed by Mokos et al. \cite{mokos2025model}. Unlike prior works focused on validation or verification, their approach emphasizes generating Semantics of the System-Level Integrated Modeling (SLIM) models from System Attributes Ontology (SAO) and Extended Attributes Ontology (EAO).
%These ontologies formalize restricted natural language, enabling reasoning over ambiguities and inconsistencies. SLIM models support execution semantics, fault modeling, and traceability through translation into state machines. Verification is achieved via Linear Temporal Logic (LTL) property derivation and analysis using COMPASS tools \cite{bozzano2019compass}. This unified workflow advances ontology-based assurance for safety-critical systems.

Mokos et al. \cite{mokos2025model} propose a semi-automated framework linking informal requirements to formal safety analysis via ontology-driven synthesis. Unlike prior works on validation or verification, their focus is on generating SLIM models from System Attributes Ontology (SAO) and Extended Attributes Ontology (EAO). These ontologies formalize restricted natural language, supporting reasoning over ambiguities and inconsistencies. SLIM models provide execution semantics, fault modeling, and traceability through state machine translation. Verification is performed by deriving Linear Temporal Logic (LTL) properties and analyzing them with COMPASS tools \cite{bozzano2019compass}. This unified workflow advances ontology-based assurance for safety-critical systems.

%Conclusion
\paragraph{\textbf{Conclusion for the Theoretical Foundations domain.}} In this domain, all contributions employ ontologies with OWL DL, but there is no evidence of KGs, semantic log, or ML/LLM integration, reflecting a purely symbolic orientation. The works focus on formal specification, analysis, and verification of software and hardware, often combining ontologies with formal reasoning approaches. For instance, some studies integrate FF or TL, making this one of the few domains with explicit attention to advanced reasoning. Ontology modeling here is split between static and dynamic approaches, with most studies delivering VS+S solutions that provide operational or system-level support, while one contribution is limited to VS. Despite these strengths, ontology reuse and openness are absent, and no progress is made toward hybridization with data-driven methods.

\definecolor{lightorange}{RGB}{255,245,230}
\begin{tcolorbox}[
  colback=lightorange,   % background color
  colframe=lightorange,  % border color
  boxrule=0pt,           % no visible border
  arc=0mm,               % sharp corners
  left=2mm, right=2mm,   % padding
  top=1mm, bottom=1mm
]

\textbf{Summary:} Table \ref{tab:domian_theoretical_foundations} shows that the Theoretical Foundations domain is marked by a strong reliance on formal methods and ontology-based reasoning, which sets it apart from other domains. Nevertheless, the lack of KGs, semantic log integration, hybrid ML/LLM approaches, and openly shared ontologies limits its applicability in data-intensive, real-world cybersecurity contexts.
\end{tcolorbox}

\section{CONCLUSION AND FUTURE WORK}
We conducted a comprehensive survey of existing ontological studies across the cybersecurity domains and subdomains defined by the JRC Cybersecurity Taxonomy. In total, 73 papers were reviewed, spanning 8 domains and 25 selected subdomains, with an average of 3 studies per subdomain. The findings highlight that advancing the use of ontologies in cybersecurity requires addressing several foundational challenges.

%One such challenge was the application of the JRC Cybersecurity Taxonomy, where we encountered significant difficulties in consistently classifying articles within its domains and subdomains. This is partly because certain  overlap or span across multiple domains or subdomains. For example, Incident Handling and Digital Forensics could reasonably be treated as two separate domains rather than a single.

In ontology research for cybersecurity, the most pressing challenge is the lack of formal validation and verification (V\&V). Although numerous ontologies have been developed, few undergo thorough evaluation, with critical aspects such as logical consistency, reasoning capabilities, and cross-context usability often overlooked. This absence of systematic assessment undermines both the reliability and practical utility of these ontologies. Informal or ad hoc evaluation remains prevalent, limiting confidence in their deployment for real-world applications. To build trust and adoption, the community must use rigorous, tool-supported V\&V, strengthening both the theory and practice of cybersecurity ontologies.

Continued research may prioritize the development of standardized and replicable V\&V procedures. This includes the formal validation of ontologies using OWL reasoners, the application of competency-question-based evaluation frameworks, and the use of SHACL (Shapes Constraint Language), a W3C standard for validating RDF graphs by checking structural conformance against defined constraints. Moreover, the creation of comprehensive evaluation checklists—derived from criteria synthesized across existing studies—could serve as practical tools for ontology developers to ensure quality and contextual fitness. Advancing these practices will help bridge the current methodological gap and foster greater confidence in ontology-driven cybersecurity solutions.

Another significant challenge is the limited accessibility of most cybersecurity ontologies. They are often developed for specific projects and remain unpublished or difficult to access. Furthermore, the field lacks a centralized repository or dedicated platform to aggregate and disseminate these ontologies. This is contrary to the FAIR principles, which emphasize that scientific resources should be: \emph{Findable}, \emph{Accessible}, \emph{Interoperable}, and \emph{Reusable}.

In contrast, domains such as bioinformatics benefit from well-established infrastructures like the Biomedical Ontology Repository (BioPortal\footnote{\url{https://bioportal.bioontology.org/}}), which embody the FAIR principles by promoting best practices in sharing, discoverability, interoperability, and reuse. The absence of a comparable infrastructure in cybersecurity undermines the adoption of FAIR practices, ultimately limiting progress in knowledge integration, semantic interoperability, and collaborative development.

A critical area in cybersecurity is the Incident Handling and Digital Forensics domain. Despite notable progress, several challenges persist. There is limited integration between semantic reasoning and efficient data-level mechanisms, affecting scalability. Support for end-to-end forensic integrity remains weak, especially under adversarial conditions using anti-forensic techniques \cite{nisioti2023forensics}. Current ontologies also lack unified modeling of the full incident response lifecycle. Other gaps include poor handling of uncertainty, limited causality tracking, and weak cross-layer reasoning across technical and strategic levels.

Neuro-symbolic approaches \cite{grov2024neurosymbolic} hold particular promise by combining ontology-based reasoning with machine learning (ML), enabling capabilities such as automated incident classification, threat correlation, and behavior prediction \cite{ghidalia2024combining}. Furthermore, cross-organizational and cross-jurisdictional ontologies are essential to support real-time, interoperable responses. Achieving effective global incident handling also requires careful alignment with heterogeneous legal and policy environments. Dynamic ontology evolution using LLMs and NLP can automate updates from logs and reports. Such tools improve adaptability to emerging threats and operational changes \cite{cotti2025enabling, vassiliadis2024challenges}. These advancements will enhance cybersecurity readiness and response effectiveness. They also bridge the gap between formal knowledge models and real-world applications.

\begin{acks}
We would like to thank all researchers at the Institute of Systems and Computer Engineering – Research and Development (INESC-ID) for their valuable contributions and constructive feedback, as well as the Portuguese Navy Research Center (CINAV) for their continuous support throughout this doctoral research. We also acknowledge the support of the SafeIaC project (LISBOA2030-FEDER-00866800). This work was further funded by national funds through the Fundação para a Ciência e a Tecnologia (FCT), under project reference UIDB/50021/2020 (http://dx.doi.org/10.54499/UIDB/50021/2020
).
\end{acks}

\bibliographystyle{ACM-Reference-Format}
\bibliography{my}

\appendix

\section{Research Methods}

\subsection{\textbf{Section 2}}
The review begins by exploring the role of ontologies as formal semantic frameworks for representing cybersecurity knowledge (\textbf{Section~\hyperref[sec:2.1]{2.1}}).
Next, \textbf{Section~\hyperref[sec:2.2]{2.2}} defines the review scope by selecting 8 cybersecurity domains from the JRC Cybersecurity Taxonomy \cite{JRC2021CybersecurityTaxonomy, Nai2019}, based on their strategic relevance, research activity, and technical depth.
\textbf{Section~\hyperref[sec:2.3]{2.3}} outlines the methodology, including systematic keyword-based searches across major academic databases, predefined inclusion criteria, and a comparative framework for ontology evaluation.

\subsection{\textbf{Section 3}}
Key prior surveys on cybersecurity ontologies are analyzed, with emphasis on their thematic focus, classification strategies, and methodological approaches.

\subsection{\textbf{Section 4}}
Existing cybersecurity ontologies are reviewed by domain and subdomain, following the structure defined by the JRC Cybersecurity Taxonomy.
A comparative analysis highlights the modeling strategies, scope, and contributions of each approach, along with the emerging roles of semantic log extraction and LLMs.

\subsection{\textbf{Section 5}}
The conclusion synthesizes findings from the reviewed studies, identifying persistent gaps in validation, reasoning capabilities, and deployment. Future work is directed toward adopting rigorous, tool-supported verification and validation methods to improve the reliability and practical use of cybersecurity ontologies.

\section{Online Resources}
Relevant studies were identified through search strings applied in scholarly databases, including Google Scholar, Web of Science, ACM Digital Library, and IEEE Xplore.

\end{document}